\documentclass[namedreferences]{SolarPhysics}
\usepackage[optionalrh]{spr-sola-addons} 
\usepackage{graphicx}        
\usepackage{color}           
\usepackage{url}             

\renewcommand{\vec}[1]{ {\mathbf #1} }

\newcommand{\aap}{    {\it Astron. Astrophys.}}

\newcommand{\aapr}{   {\it Astron. Astrophys. Rev.}}

\newcommand{\apj}{    {\it Astrophys. J.}}

\newcommand{\grl}{    {\it Geophys. Res. Lett.}}

\newcommand{\jgr}{    {\it J. Geophys. Res.}}

\newcommand{\pre}{    {\it Phys. Rev. E}}
\newcommand{\solphys}{{\it Solar Phys.}}

\begin{document}

\begin{article}

\begin{opening}

\title{Vlasov-Maxwell, self-consistent electromagnetic wave emission simulations in the solar corona}

\author{David~\surname{Tsiklauri}$^{1}$     
       }
\runningauthor{Tsiklauri}
\runningtitle{Vlasov-Maxwell simulations of radio emission in solar corona}

   \institute{$^{1}$Astronomy Unit, School of Mathematical Sciences, Queen Mary University of London, Mile End Road, London, E1 4NS, United Kingdom}

\begin{abstract}
1.5D Vlasov-Maxwell simulations are employed to model electromagnetic emission generation in a fully 
self-consistent plasma kinetic model for the first time in the solar physics context. The simulations mimic the plasma emission mechanism and Larmor drift instability
in a plasma thread that connects the Sun to Earth with 
the spatial scales compressed appropriately. The effects of spatial density gradients on the generation 
of electromagnetic radiation are  investigated. 
It is shown that 1.5D inhomogeneous plasma with a uniform background 
magnetic field directed transverse to the density
gradient is aperiodically unstable to Larmor-drift instability. The latter
results in a novel effect of generation of electromagnetic emission at plasma frequency. 
The generated perturbations
consist of two parts: (i) non-escaping (trapped) Langmuir type oscillations which are
localised in the regions of density inhomogeneity, and are highly filamentary, with the period of 
appearance of the filaments close to
electron plasma frequency in the dense regions; and (ii) escaping electromagnetic radiation with phase 
speeds close to the
speed of light. When density gradient is removed (i.e. when plasma becomes stable to Larmor-drift instability)
and a {\it low density} super-thermal, hot beam is injected along the domain, in the direction perpendicular to the magnetic field,
plasma emission mechanism generates non-escaping Langmuir type oscillations
which in turn generate escaping electromagnetic radiation.
It is found that in the spatial location where the beam is injected, the standing waves, 
oscillating at the plasma frequency, are excited. These can be used to interpret
the horizontal strips (the narrowband line emission) observed in some dynamical spectra.
Quasilinear theory predictions: (i) the electron free streaming and (ii) the beam long relaxation
time, in accord with the analytic expressions, are corroborated via direct,
fully-kinetic simulation.
Finally, the interplay of Larmor-drift instability and plasma emission mechanism is studied
by considering {\it dense} electron beam in the Larmor-drift unstable (inhomogeneous) plasma. 
The latter case enables one  to study the deviations from the quasilinear theory.
\end{abstract}
\end{opening}

\section{Introduction}
The aim of this paper is to employ Vlasov-Maxwell simulations for the electromagnetic wave generation by a 
super-thermal ($0.2-0.5c$), hot electron beam injected into the solar coronal magnetised plasma. 
Since such beams are thought to be responsible for the generation of type III solar radio 
bursts we start from a brief review of the previous relevant results. Although we stress 
that for the reasons described below, at this stage, the model presented here cannot be 
directly applied to the type III busts and it is perhaps more relevant for the 
interpretation of the narrowband line emission observations.      
The type III solar radio bursts are believed to come from the
super-thermal beam of electrons that travel away from the Sun
producing the observed electromagnetic (EM) radiation via the plasma
emission mechanism (see e.g. \opencite{1987SoPh..111...89M,2008SoPh..253....3N,2008A&ARv..16....1P} for recent reviews).
Basic physical understanding of the generation of type III radio burst 
EM waves in plasma by an electron beam has been with us for
over five decades \cite{1958SvA.....2..653G} and involves the generation of Langmuir waves by bump-on-tail unstable
electron distributions and subsequent mode conversion of the longitudinal Langmuir waves into
escaping, transverse EM radiation.
Theoretical efforts in understanding of type III solar radio bursts can be grouped into three categories:

(1) Quasilinear theory of type III sources that use kinetic Fokker-Planck type equation for describing
the dynamics of an electron beam, coupled with spectral energy density evolutionary equations for  
Langmuir and ion-sound waves have long been
studied. The most essential result is that the spectral energy density of the Langmuir wave packets (that are excited by the
bump-on-tail unstable beam) travels along
the open magnetic field lines with a constant speed and this is despite the quasilinear relaxation 
(formation of a plateau in the longitudinal, along the beam injection direction, velocity 
phase-space of the electron distribution function), hence, this implies some sort of beam marginal
stabilisation \cite{1968SvA....11..956K,1970AdA&A...7..147S,1972SoPh...24..444Z,1999SoPh..184..353M,2002PhRvE..65f6408K}.
Inclusion of EM emission component into the quasilinear theory in some models is
based on so-called 
drift approximation  
\cite{1988A&A...195..301H,1990A&A...229..216H,1999A&A...342..271H},
where nonlinear beam stabilisation during its propagation (so called free streaming) is based on Langmuir-ion acoustic wave coupling
via ponder-motive force and EM emission is prescribed by a power law  of the beam to ambient plasma number density ratio.
These models can be successfully compared with the observed dynamical spectra to constrain key model
parameters. 

(2) Stochastic growth theory \cite{1992SoPh..139..147R,1992ApJ...387L.101R}, in which density irregularities induce
random growth, such that Langmuir waves are generated stochastically
and quasilinear interactions within these Langmuir
clumps cause the beam to fluctuate about marginal stability. Further this approach has been developed into
a numerical simulation tool that effectively can reproduce observational features of the type III bursts \cite{2008JGRA..11306104L}.

(3) Full kinetic simulation 
approach of type III bursts \cite{2001JGR...10618693K,2005ApJ...622L.157S,2009ApJ...694..618R,2010JGRA..11501204U}  to this 
date used Particle-In-Cell (PIC) numerical method. This effort is mainly focused on understanding of basic physics rather than
direct comparison with the observations because the size of simulation domain of the models corresponds to only few 1000 Debye lengths which
is roughly $1 / 10^{10}$th of 1 AU. Thus, such models deal with micro-scales.

Each of the above theoretical approaches have their advantages and disadvantages.
For example as it is well explained in \inlinecite{Cairns85}, in quasilinear theory the particle distribution
function is split into a slowly varying average part and a rapidly varying part due to a wave.
Then major approximation is that interactions between wave modes are neglected and only
back-reaction of the waves on the  slowly varying average part of the distribution function is considered.
Therefore, if computational resources permit, the full kinetic approach is more desirable. 
However, to this date only PIC method was used. It is well known that PIC approach suffers from large, so called,
"shot-noise" level of which scales as one over the root of number of particles. Moreover, in PIC approach typically
there are few hundred particles per cell (which is normally one Debye length long in fully electromagnetic PIC codes).
In Vlasov-Maxwell approach instead of solving for individual particle dynamics, without loss of
generality (or any kinetic physics), collisionless Vlasov equation for electrons and ions is solved in which
EM fields are self-consistent. In this study we use 80$\times$80 velocity grid which in PIC
equivalent would be having $80\times80=6400$ particles per cell (instead of few hundred). We have also done
convergence tests by increasing the resolution both in velocity space and spatial domain and confirmed 
the results' convergence. This shows that Vlasov-Maxwell approach probes more finely phase space of the problem
but this comes at a substantial memory cost.   

The paper is organised as following. In Section 2 we present the model and main results,
including (i) a numerical run of the inhomogeneous plasma without a beam which turns out to be aperiodically 
unstable to a Larmor drift. The latter results in Langmuir and EM wave generation, as well as
density filamentation (Section 2.1); (ii) a numerical run of the homogeneous plasma with a low density beam
which generates  Langmuir and EM wave via plasma emission mechanism (Section 2.2); and
 a numerical run of the inhomogeneous plasma with a high density beam in order to combine
 the effects of both density inhomogeneity and the presence of the beam  (Section 2.3).
 
 \section{The model and general theoretical considerations}
 
Our numerical model is implemented using a relativistic, fully electromagnetic 
Vlasov-Maxwell code called VALIS \cite{2009JCoPh.228.4773S}. The code is using
a conservative, split-Eulerian scheme based on the Piecewise Parabolic Method for the update of 
the particle distribution function and utilising the exact particle fluxes to 
calculate the current in the solution of Maxwell's equations.
In particular relativistic Vlasov's equation is solved for species $\alpha$ ($\alpha=e,i$ for 
electrons and ions respectively)
\begin{equation}
\frac{\partial f_\alpha}{\partial t} + \frac{\vec u}{\gamma^*} \cdot \vec \nabla_x f_\alpha
+\frac{q_\alpha}{m_\alpha}\left(\vec E + \frac{\vec u}{\gamma^*}\times \vec B\right) \cdot 
\vec \nabla_{\vec u} f_\alpha=0
\end{equation} 
in conjunction with the Maxwell's equations 
\begin{equation}
\vec \nabla \cdot \vec E=\rho/\varepsilon_0, \;\;\;\; \vec \nabla \cdot \vec B=0,
\;\;\;\; \vec \nabla \times \vec E = -\frac{\partial \vec B}{\partial t}, 
\;\;\;\;\vec \nabla \times \vec B = \mu_0 \vec J + \frac{1}{c^2} \frac{\partial \vec E}{\partial t},
\end{equation} 
where charge and current densities are specified in {\it self-consistent} manner,
\begin{equation}
\rho=\sum_\alpha q_\alpha \int f_\alpha d^3u, \;\;\;\;\;
\vec J=\sum_\alpha q_\alpha \int (\vec u /\gamma^*) f_\alpha  d^3 u,
\end{equation}
and $\gamma^*=\sqrt{1+|\vec u|^2/c^2}$.

VALIS is 2D2V code in that it has two spatial dimensions $(x,y)$ and two corresponding
velocity components $(u_x,u_y)$, while electric and magnetic field components that are solved for
are $(E_x,E_y,0)$ and $(0,0,B_z)$. Distance and time are normalised to $c / \omega_{pe}$ and $\omega_{pe}^{-1}$, while
electric and magnetic fields to $\omega_{pe}c m_e /e$ and  $\omega_{pe} m_e /e$ respectively. 
Temperature is normalised to to $m_e c^2/k$. Here $\omega_{pe} = \sqrt{n_e e^2/(\varepsilon_0 m_e)}$ is the electron plasma frequency,
$n_\alpha=\int f_\alpha d^3u$ is the number density and all other symbols have their usual meaning.

We intend to consider a single plasma thread (i.e. to use 1.5D geometry), therefore
space components considered are $(x,y)=(25000 \lambda_D, 1 \lambda_D)$ with $\lambda_D = v_{th,e}/ \omega_{pe}$
being Debye length (here $v_{th,e}=\sqrt{k T/m_e}$ is electron thermal speed).
We would like to resolve full plasma kinetics, 
therefore we set spatial grid size as $1 \lambda_D$. In practice this means
we set plasma temperature at $T=10^5$K (i.e. fix  $v_{th,e}$ at $4.12\times10^{-3} c$),
which corresponds to high solar corona, above active regions.
In the presented results (and in the VALIS code generally) spatial scales are normalised to $c/\omega_{pe}$.
We do not fix  plasma
number density and hence $\omega_{pe}$ deliberately, because we wish our results to stay general.
In order to achieve this generality (and consistency of the results) it 
is important to keep normalised $\tilde B_{z0}=  B_{z0} / (\omega_{pe} m_e /e)= 0.01$ the same in all numerical runs.
Because  $B_{z0}$ is normalised to $\omega_{pe} m_e /e$, no matter how plasma
density and hence $\omega_{pe}$ changes, (i) ratio of Debye length and $c/\omega_{pe}$, i.e.
$\lambda_D / (c/\omega_{pe}) = 4.12\times10^{-3}$ and 
ratio of electron Larmor radius and $c/\omega_{pe}$, i.e. $r_{L,e} / (c/\omega_{pe}) = 4.12\times10^{-1}$
{\it stay the same}. Such choice means, of course, that 
magnetic field in Tesla is variable.
For example, if we set plasma number density to
 $n_0 =10^{15}$ m$^{-3}$ (i.e. fix $\omega_{pe} = 1.78 \times 10^9$ Hz radian),
this sets Debye length at $\lambda_D = 6.90\times10^{-4}$m=$4.12\times10^{-3} c / \omega_{pe}$ 
and electron Larmor radius at $r_{L,e} = 6.90\times10^{-2}$m=$4.12\times10^{-1} c / \omega_{pe}$.
Also,  then $B_{z0} =1.01\times10^{-4}$T $\approx 1$ gauss.
If we set plasma number density to
 $n_0 =10^{-5}$ m$^{-3}$,
this sets Debye length at $\lambda_D = 6.90\times10^{6}$m=$4.12\times10^{-3} c / \omega_{pe}$ 
and electron Larmor radius at $r_{L,e} = 6.90\times10^{8}$m=$4.12\times10^{-1} c / \omega_{pe}$.
Also,  then $B_{z0} =1.01\times10^{-14}$ Tesla.
In other words, appropriately adjusting plasma number density
$n_0$, physical domain can have arbitrary size e.g. Sun-earth distance (but then
unrealistically low density has to be assumed). 
Since the number of grid points and
domain size (normalised to $c / \omega_{pe}$) is set independently,
we have to make sure that grid size is $1 \lambda_D$ by setting $n_x=25000$ and $L_{x,max} = 25000\times \lambda_D=
102.94 c / \omega_{pe}$ and $n_y=1$ and $L_{y,max} = 1\times \lambda_D= 4.12\times10^{-3} c / \omega_{pe}$.
For the velocity we have $(u_x,u_y)=(80,80)$ grid points with maximal possible velocities for electrons
allowed set $u_{x,max}=u_{y,max} = 0.25c$ ($u_{x,max}=u_{y,max} = 0.4c$ in Section 2.3) 
for electrons and $u_{x,max}=u_{y,max} = 0.25/\sqrt{1836}c=5.83\times10^{-3}c$ for ions.
Since the code does not allow to set magnetic field along $x$-axis (because $\vec B=(0,0,B_z)$), 
to represent the situation adequately, we have only a choice
to set a $B_{z0}$ component, which we fix at 0.01 in normalised units. 
This is not an unreasonable value for, transverse to the considered plasma thread, component of magnetic field above an active region.
At first, it seems unrealistic to ignore magnetic field along $x$. However, firstly, bulk of the work 
in the quasilinear theory indeed makes the same assumption (ignores longitudinal magnetic field).
Secondly, it is known \cite{abr88} that in the case of weak fields, general picture of excitation of the Langmuir waves
by a low density ($n_b \ll n_e$) electron beam (i.e. their resonant interaction) via Cherenkov resonance is not much
different from the case without the magnetic field. 

In all presented numerical runs boundary conditions for all quantities are periodic.
A typical numerical run takes 32 hours on 256 processor cores (Dual Quad-core Xeon, eight cores per computing node).

We now briefly re-iterate key facts about beam-plasma interaction theoretical framework.
In the case without magnetic field (or in the weak field case) cold plasma 
dispersion relation yields
two possible modes (e.g. \opencite{abr88}, p.156):
\begin{equation}
\omega^2=k^2c^2+\omega_{pe}^2+\omega_{pb}^2\gamma^{-1},
\end{equation} 
\begin{equation}
1-\frac{\omega_{pe}^2}{\omega^2}-\frac{\omega_{pb}^2\gamma^{-3}}{(\omega-k_\parallel v_b)^2}
\left[1+ \frac{k_\perp^2 v_b^2\gamma^2}{\omega^2}\right]=0.
\end{equation} 
Here, $\omega_{pb}=\sqrt{n_b e^2/(\varepsilon_0 m_e)}$ the beam plasma frequency and $v_b$ is
its speed. $\gamma$ is the usual Lorentz factor for the bulk motion of the beam
(in cold plasma approximation thermal motions of plasma are absent). 
Note that here only electron and beam contributions are retained whilst
ion contribution is ignored due to its smallness.
Equation (4) describes a stable, purely transverse ($\vec E \perp \vec k$), EM wave which does not interact 
with the beam because $\vec E \cdot \vec v_b=0$. If $B_0\parallel z$ then in the {\it cold} plasma
approximation beam can only propagate along $z$-axis
(Note that our numerical simulations are with {\it finite} temperature). EM wave described by Equation (4) has non zero $E_y$ component only. 
Equation (5) describes oblique wave which has both $E_\parallel=E_z$ and $E_x$ components and hence
can interact with the beam via Cherenkov resonance. By putting $\omega=k_\parallel v_b+\delta = k_z v_b+\delta$
into Equation (5) growth rates (for $k_\parallel v_b \leq \omega_{pe}$ when the oblique
mode becomes unstable), $\delta$, can be easily found:
(i) away from the plasma frequency, $\omega^2\approx k_\parallel^2 v_b^2 \not =\omega_{pe}^2$ (non-resonant case),
\begin{equation}
\delta_{nr}=\frac{\omega_{pb}\gamma^{-3/2}}{\sqrt{1-\omega_{pe}^2/(k_\parallel^2 v_b^2)}}\left[\frac{k_\parallel^2+k_\perp^2\gamma^2}
{k_\parallel^2}\right]^{1/2}, \;\;\;
Im(\delta_{nr})\propto \omega_{pe}\left(\frac{n_b}{n_e}\right)^{1/2}
\end{equation}
(ii) close to the plasma frequency, $\omega^2\approx k_\parallel^2 v_b^2 \approx \omega_{pe}^2$ (resonant case),
\begin{equation}
\delta_r= \omega_{pe}\left[\frac{n_b}{2 n_e} \frac{1}{\gamma^3} 
\frac{k_\parallel^2+k_\perp^2\gamma^2}
{k_\parallel^2}\right]^{1/3}, \;\;\;
 Im(\delta_r) \propto \omega_{pe}\left(\frac{n_b}{n_e}\right)^{1/3}.
\end{equation}
Naturally, the resonant growth rate is much larger than the non-resonant one, 
$\delta_{r} \gg \delta_{nr}$, due to the beam's low density ($n_b \ll n_e$).
As is also clear from Equation (5), the Cherenkov resonance effectively excites
Langmuir (longitudinal) waves with the dispersion relation $\omega^2 \approx\omega_{pe}^2$
(note that thermal effects are ignored in Equations (4)-(9)).

In the case of strong
magnetic field ($\omega_{ce}\gg\omega_{pe}$) there are two types of waves,
dispersion relations of which are given by \cite{abr88}
\begin{equation}
k^2-\omega^2/c^2=0,
\end{equation} 
\begin{equation}
k_\perp^2+\left(k_\parallel^2-\omega^2/c^2)(1-\omega_{pe}^2/\omega^2-\omega_{pb}^2\gamma^{-3}/(\omega-k_\parallel v_b)^2\right)=0.
\end{equation} 
Our Equation (9) without the beam contribution term is also derived by \opencite{1986ApJ...302..120A}, 
their Equation (49), and is referred to as O-mode.
The beam does not interact with the purely transverse EM wave given by Equation (8), while it can
interact with the slow wave $\omega_-$ given by Equation (9) (Equation (9) describes oblique EM waves which have both $E_\parallel$
and $E_x$ and in the absence of the beam reduces to the fast ($\omega_+$) and slow ($\omega_-$) modes with dispersion \\
$\omega_{\pm}^2=0.5\left[\omega_{pe}^2+k^2c^2\pm \sqrt{(\omega_{pe}^2+k^2c^2)^2-4 
\omega_{pe}^2k_\parallel^2c^2}\right]$).
Electron beam then can resonantly interact with the slow mode 
when $\omega_-$ intersects with $\omega=k_\parallel v_b$ line.
Thus, the presence of the longitudinal magnetic field can only alter 
value of $k_\parallel$ at which the Cherenkov resonance occurs.
It should be stressed that Equations (4)-(9) are derived in the case electron beam propagating strictly 
along the magnetic field. Therefore, beam is strictly decoupled from the purely transverse EM wave (Equation (8)).
If the {\it beam} has a small $k_\perp$ 
(note that $k_\perp$ in Equations (4-9) refers to that of a wave mode)
then it can couple to EM wave and hence generate it.
This is our main motivation for have small $B_{z0}$ so that when beam is injected along $x$-axis it can couple to
the EM mode.
Also, \inlinecite{2010PhPl...17c2104H}
showed, using relativistic Vlasov equation,  that in unmagnetised plasma, 
EM and plasma wave conversion efficiency diminishes to zero
at both $0^\circ$ and $90^\circ$ incidence angles and peaks between
$10-20^\circ$ depending on plasma temperature. Here degrees refer to angle
between wavenumber $\vec k$ and density gradient direction.
Resuming aforesaid we acknowledge that the present numerical model is 
not suitable for
describing the type III solar bursts directly. Ideally, 
to represent the true physical reality, it would be 
preferable to set large $B_{x0}$ in addition to
small $B_{z0}$ (as in solar wind Parker spiral in the upper solar corona). 
However, since the VALIS code only solves for
$(E_x,E_y,0)$ and $(0,0,B_z)$ we have to simply make sure that
when the beam is injected along $x$-axis it can couple to
the EM mode in order to capture the essential physics.
At the same time we stress that existence of $B_{x0}$ is not
a requirement for the generation of type III bursts {\it per se}, what is 
essential is to have finite $k_\perp$ in the beam so that it couples to 
the EM wave (here we achieve this by setting small $B_{z0}$ only).
We also note that it was our intension to consider magnetised plasma with the 
beam injected strictly along the physical domain. In principle, the coupling to 
EM wave could have been also achieved by switching off the magnetic field altogether 
and in addition to $u_{0x}$, 
setting $u_{0y} = 0.2-0.5c$. This would have created a situation with 
non-zero $k_\perp$ too, thus facilitating the coupling of the beam to an EM wave.

VALIS code allows to set any desired number of plasma particle species.
Therefore, because we intend to study spatially localised electron beam on top
of the inhomogeneous or homogeneous Maxwellian electron-ion plasma,
we solve for three plasma species electrons, ions and the electron beam.
The dynamics of the three species, which all interact via EM interaction,
can be tracked independently in the numerical code.
Velocity distribution function for electrons and ions is always set to 
\begin{equation} 
f_{e,i}(u_x,u_y)=e^{-m_{r,e,i}(u_x^2+u_y^2)/(2 T) },
\end{equation}
where $m_{r,e} =1$ for electrons and $m_{r,i}=1836$ for ions.
When cases with the beam are considered we set the following distribution
\begin{equation}
f_{b}(u_x,u_y)= {\tilde n_b} e^{-((u_x-0.2c)^2+u_y^2)/(2 T_b) }.
\end{equation}
where ${\tilde n_b}$ is normalised beam number density (${\tilde n_b}=n_b/n_{e0}$ ) and it is 
${\tilde n_b}=5 \times 10^{-6}$ for low density beam (Section 2.2)  
and ${\tilde n_b}=5 \times 10^{-2}$ for the dense beam (Section 2.3).
The normalised number density of the background plasma in the homogeneous case (Section 2.2) is ${n_{0}}=1$.
Thermal spread of the electron beam is specified by setting $T_b=9 T= 9.0\times10^5$K.
Note that the beam is injected along the $x$-axis, transverse to the background magnetic field $B_{z0}$.
Physics of the initiation of the beam is believed to be related to the magnetic
reconnection. In 2D case reconnection electric field at a magnetic null is in 
direction out-of-plane where magnetic field lies. Therefore, it is not unreasonable to consider
situation when beam is injected as in our model. Also, beam injection transverse to the
magnetic field can result from accelerated electrons from the collapsing magnetic traps 
\cite{2004A&A...419.1159K}.
In the inhomogeneous cases (Sections 2.1 and 2.3) background plasma 
normalised number density is set to
\begin{equation}
n_0(x)= 1/ \left[ 1+10^8 e^{-[ ({x-L_{x,max}/2})/{21} ]^4 }\right]
\end{equation}
This density profile mimics a factor of $10^8$ density drop from the corona  $n_0 =10^{15}$ m$^{-3}$
to $n_{AU} =10^{7}$ m$^{-3}$ at 1 AU. Because it is known that numerically 
most precisely implementable boundary conditions are periodic ones this density profile
effectively represents mirror-periodic situation when the domain size is doubled, i.e.
at $n_0(x=0)=n_0(x=L_{x,max})=1$ while  $n_0(x=L_{x,max}/2)=10^{-8}$.
This way "useful" or "working" part of the simulation domain is 
$0\leq x \leq L_{x,max}/2$.
Spatial width of the density gradient is $L_{IH}\approx 5 c/\omega_{pe}$ (see e.g. Figure 7(c), 
thick solid curve for $90 <x <95$).
When cases with the beam are considered we set its following 
density profile:
\begin{equation}
n_b(x)= {\tilde n_b}e^{-[ ({x-5})/{3} ]^4 }
\end{equation}
which means that the beam is injected at $x=5 c/\omega_{pe}$ and its 
full width at half maximum (FWHM) is also $\approx 5 c/\omega_{pe}$ (see Figure 5(c) dotted curve).
Plasma beta in this study, based on the above parameters, is set 
to $\beta=c_s^2/v_A^2=v_{th,i}^2/v_A^2 = (v_{th,i}/c)^2(\omega_{pi}/\omega_{ci})^2 =0.17$. 
($c_s$ and $v_A$ are sound and Alfven speeds respectively.)
It should be noted that, strictly speaking, pressure balance in the initial conditions
is not kept. There are two reasons for this: (i) solar wind is not in "pressure balance" 
and it is a continually expanding solar atmosphere solution; (ii) plasma beta is small
therefore it is not crucial to keep {\it thermodynamic} pressure in balance
(because its effect on total balance is negligible) and
the initial background density stays intact throughout the simulation time (see e.g. Figure 7(c), 
thick solid curve for $90 <x <95$).

\subsection{Larmor drift-unstable case, inhomogeneous plasma without a beam}

It is well known that the mode conversion from electrostatic to EM 
waves near the plasma frequency is possible by linear coupling on a density gradient.
\inlinecite{1998GeoRL..25.2609Y} examined the mode conversion from electrostatic to EM 
waves near the plasma frequency in the Earth's electron fore-shock. 
The conversion and reflection coefficients were obtained by solving coupled 
differential equations in a weakly magnetised warm plasma with a longitudinal 
linear density gradient. Results indicated that the fore-shock first harmonic 
EM emissions and the backward-propagating Langmuir waves required 
for the generation of the second harmonic EM waves could be efficiently 
generated by the linear conversion process in an inhomogeneous plasma. 
Therefore, originally the aim was to study super-thermal beam injection into 
plasma with homogeneous and inhomogeneous plasma to study the effect of the
density gradient on the level of EM wave generation.
However, we found originally unforeseen outcome in that with or without electron beam
background density gradient regions generate perturbations in all quantities 
$n(x)$, $E_x$, $E_y$ and $B_z$. The results are presented in time-distance
plots in Figure (1) and Figures (2)-(3). Dynamical picture is presented in movie 1 in the electronic supplement to this article.

\begin{figure}    
   \centerline{\hspace*{0.015\textwidth}
               \includegraphics[width=0.515\textwidth,clip=]{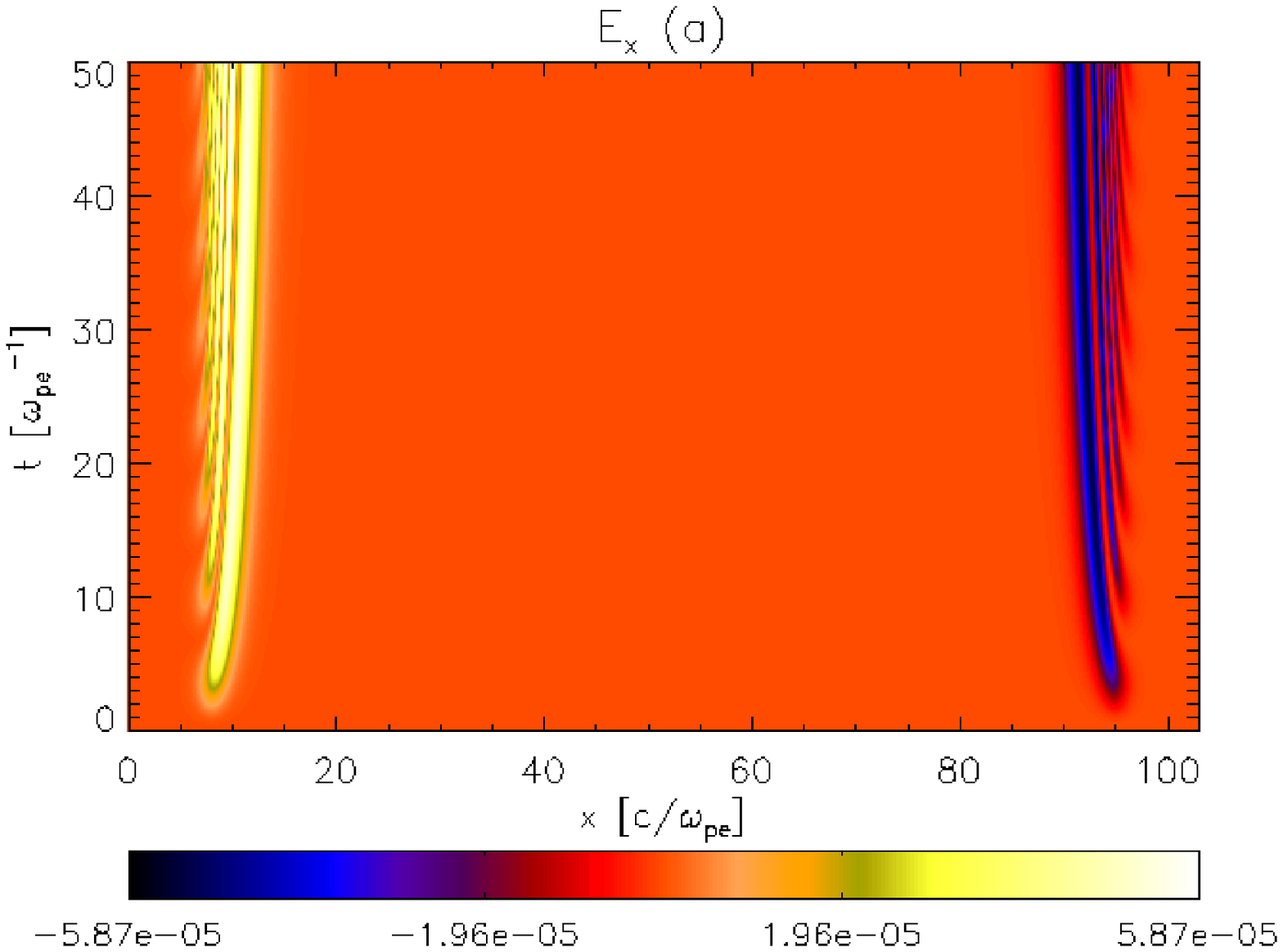}
               \hspace*{-0.03\textwidth}
               \includegraphics[width=0.515\textwidth,clip=]{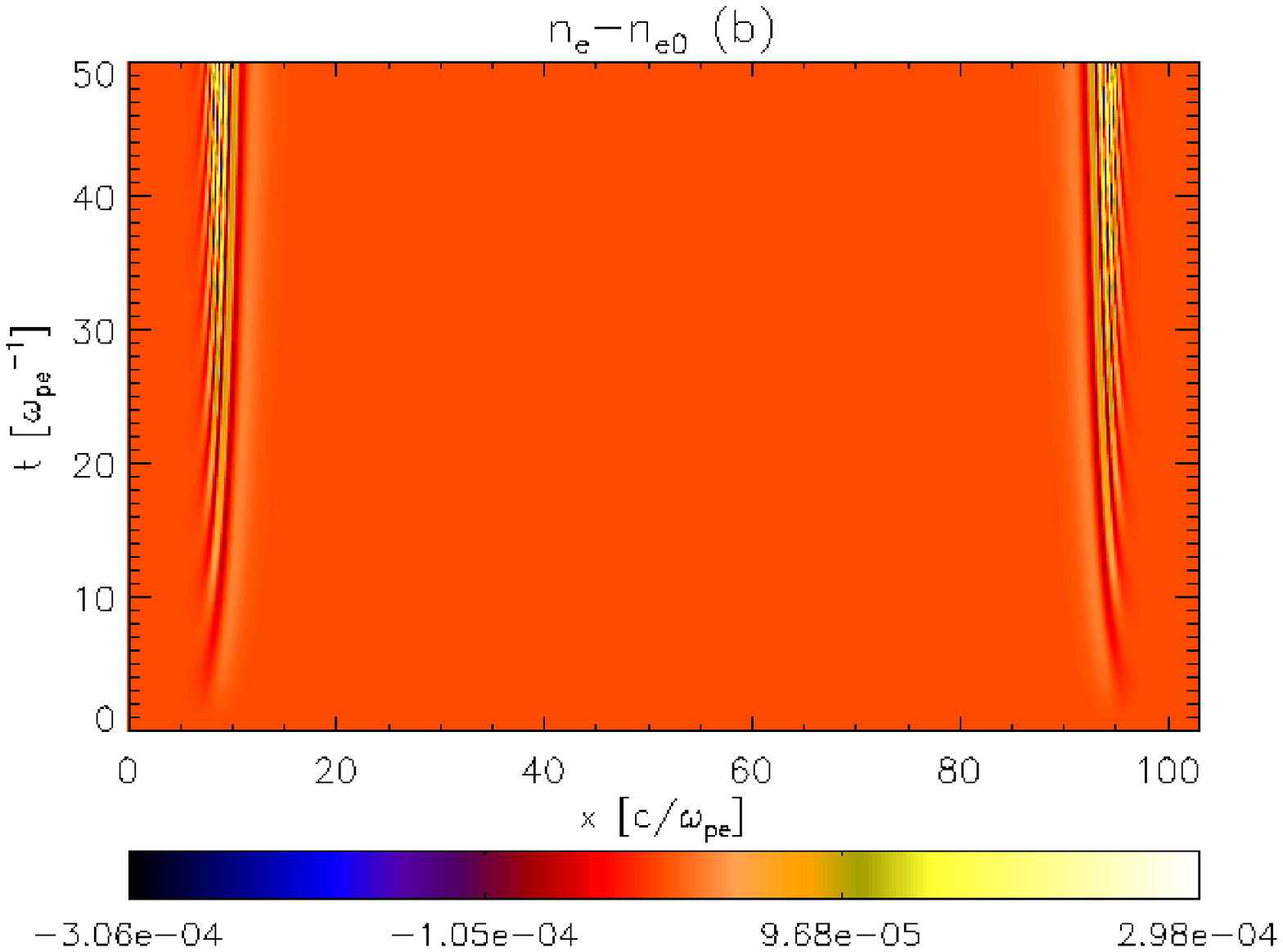}
              }
   \centerline{\hspace*{0.015\textwidth}
               \includegraphics[width=0.515\textwidth,clip=]{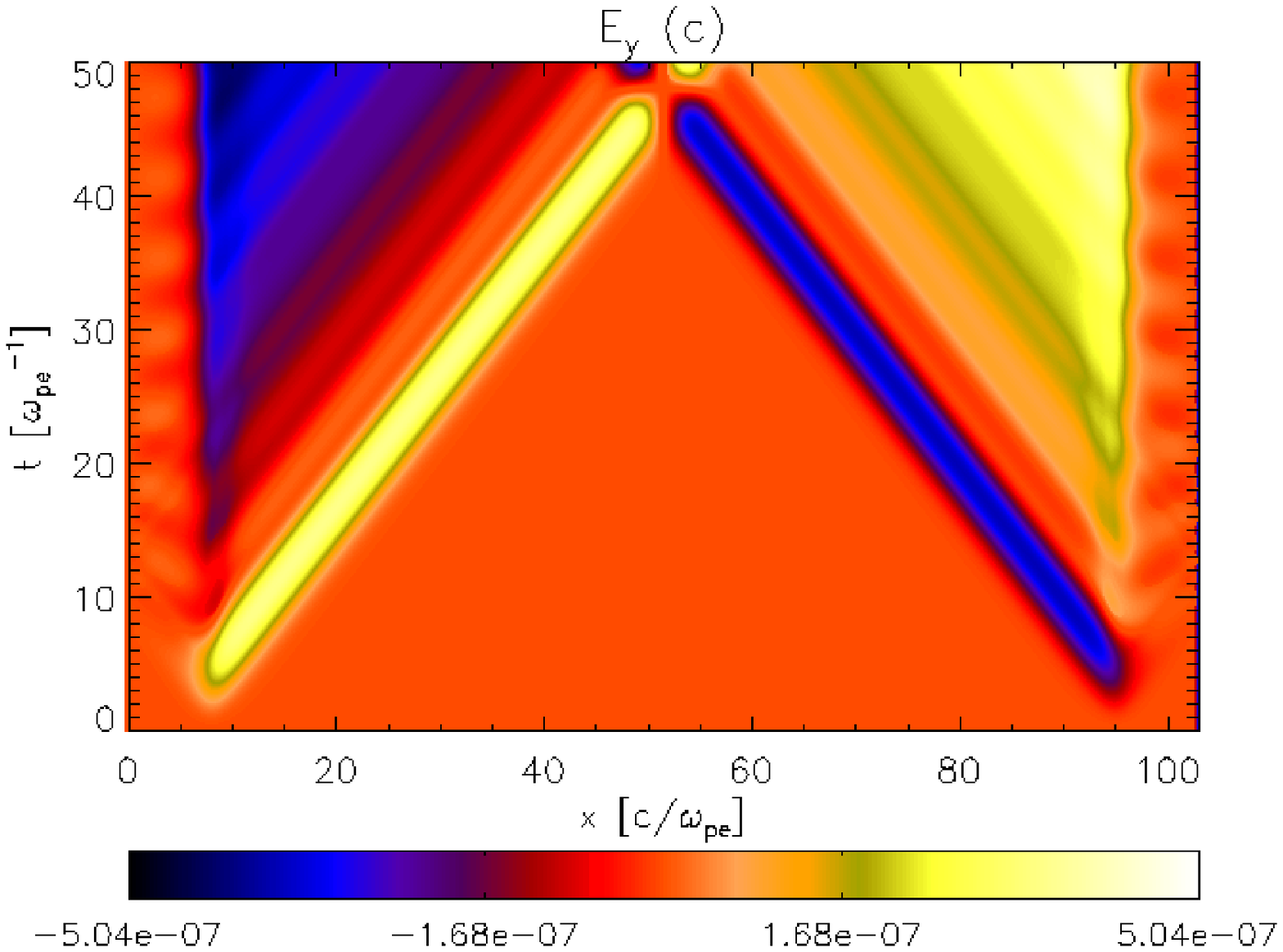}
               \hspace*{-0.03\textwidth}
               \includegraphics[width=0.515\textwidth,clip=]{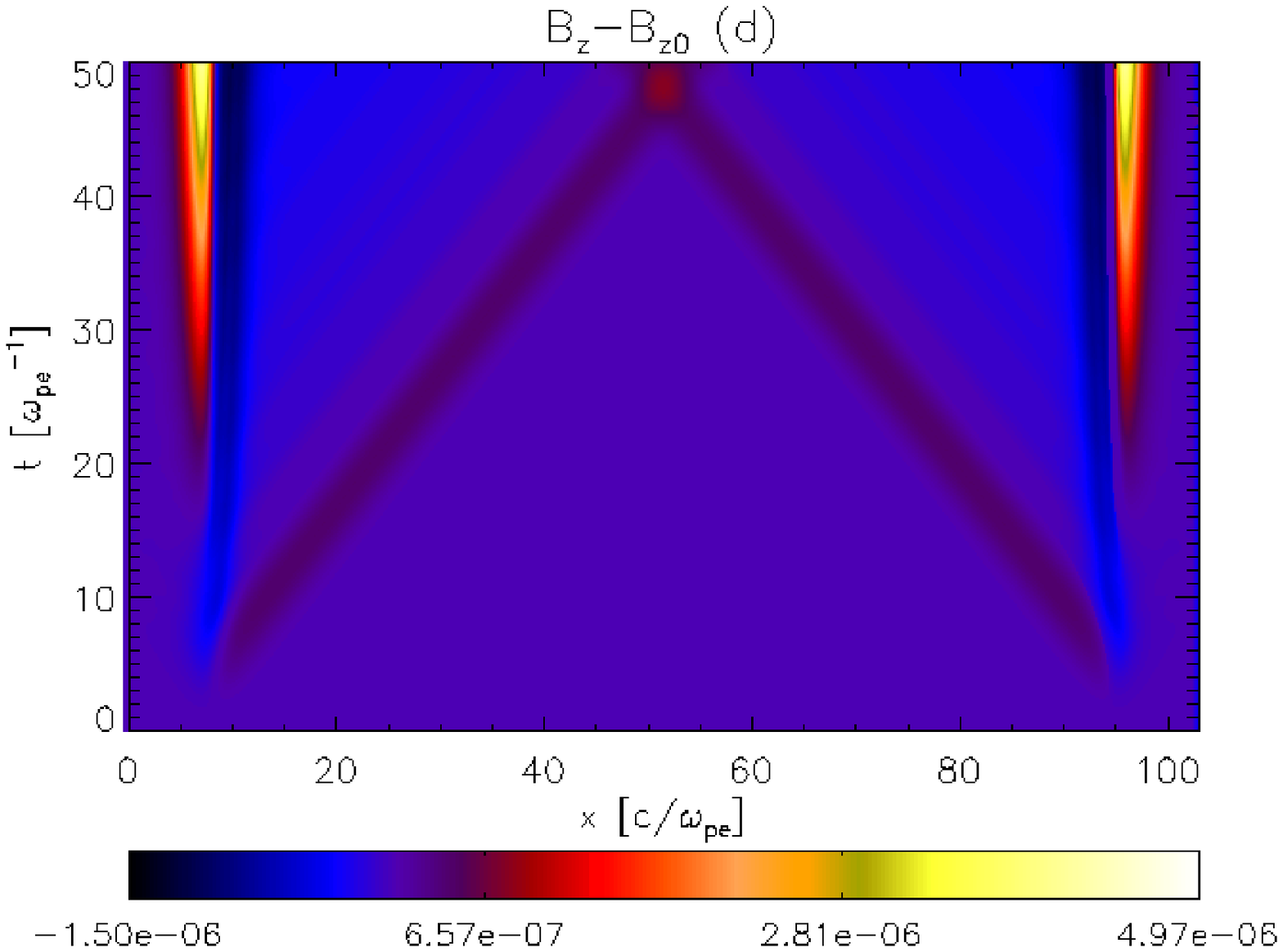}
              }
\caption{Time-distance plots for: (a) $E_x$,  (b) $n_e -n_{e0}$, (c) $E_y$ 
and  (d) $B_z-B_{z0}$.}
   \end{figure}
   
 \begin{figure}    
   \centerline{\includegraphics[width=0.99\textwidth,clip=]{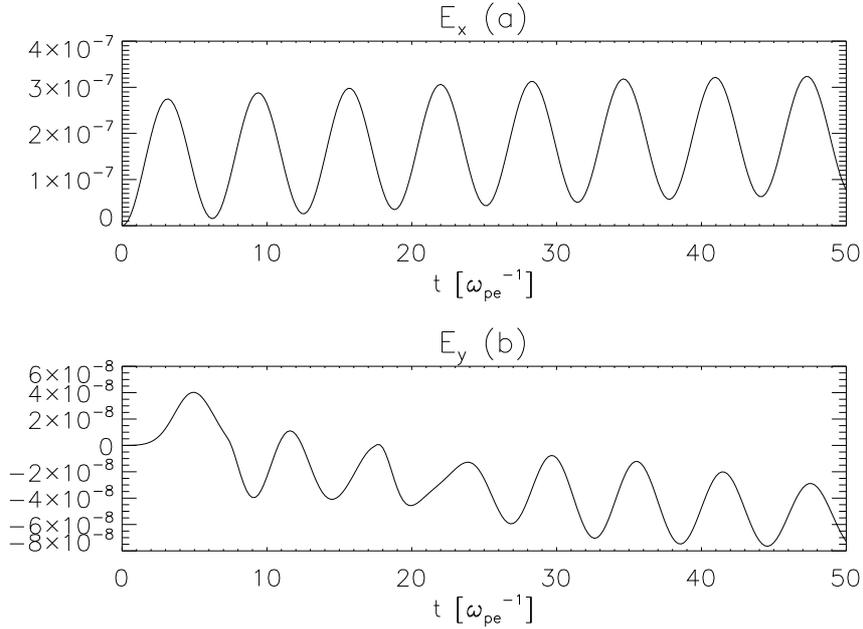}
              }
              \caption{Time evolution of: (a) $E_x(x=5,t)$ and (b) $E_y(x=5,t)$.}
      \end{figure}
   
   \begin{figure}    
   \centerline{\includegraphics[width=0.99\textwidth,clip=]{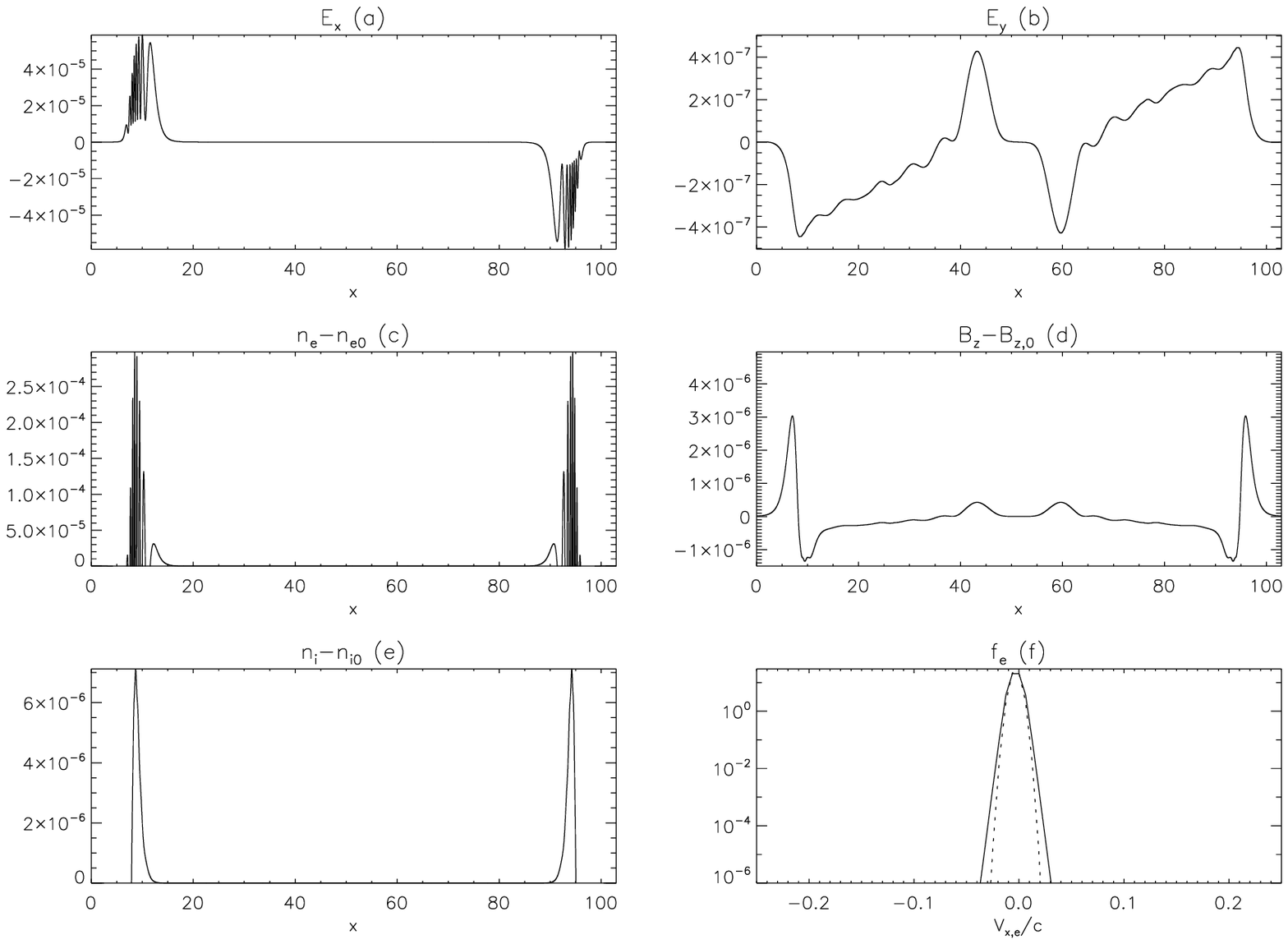}
              }
              \caption{(a) $E_x(x,t=50)$, 
 (b) $E_y(x,t=40)$,  (c) $n_{e}(x,t=50)-n_{e0}$,  (d) $B_z(x,t=40)-B_{z0}$, 
 (e) $n_{i}(x,t=50)-n_{i0}$,  (f) $f_e(v_x)$.
In (f) dotted curve represents $f_e(v_x,t=0)$ while solid one $f_e(v_x,t=50)$.
For $E_y$ and $B_z-B_{z0}$ the time snapshot $t=40$ is chosen before the end of the
simulation time, $t=50$, in order to show the spatial profiles before EM fronts collide.
It can be seen in Figure 1(c) (look horizontally across $t\approx50$) that  EM fronts
collide at about  $t\approx 50$.
}
     \end{figure}

As shown below, the obtained results can be interpreted by means of Larmor-drift unstable mode
\opencite{abr88}, p.239. 
Therefore before discussing this numerical run results, let us
briefly summarise key facts about the Larmor-drift instability.
For frequencies smaller that the Larmor-drift frequency
$\omega \leq \omega_{LD}=k_y v_{th,\alpha}^2/(\omega_{c\alpha} L_{IH})$  this mode is aperiodically
unstable in certain regimes (which as we will see below are always 
taking place in our model). Physical meaning of Larmor-drift in inhomogeneous plasma is clear.
When magnetic field is directed along $z$ and plasma inhomogeneity is along $x$-axis,
variation of the particle Larmor radii (due to the inhomogeneity) generates transverse to the both directions
current $J_y \approx q_\alpha n_\alpha v_{th,\alpha}^2/(\omega_{c\alpha} L_{IH})$.
This drift is {\it not} related to the actual motion of centres of the Larmor orbits, and it
is preferentially realised in low beta plasmas when particles are magnetised.
It is important to note that such Larmor-drift instabilities may occur in
Maxwellian plasmas and they are not related to the existence of a positive-sloped region
in the velocity distribution function. In this sense the instability can be regarded as
hydrodynamic. In the limit of long wavelength approximation, $\lambda_\perp \gg r_{L,i}$, and $\omega \gg k_z v_{th,e}$,
(which naturally holds because in our case $k_z \to 0$ because our domain size is 
infinite in $z$-direction); also when $\omega \ll \omega_{LD}$; $\omega \ll \omega_{ci}$; and $\omega_{pi} \gg \omega_{ci}$
(for our choice of parameters $\omega_{pi} / \omega_{ci}= c/v_A=4.29\times 10^3 >>1$ and this holds for arbitrary
background plasma number density),
\inlinecite{1964SvPhU...7..209R} derived following dispersion relation for the
Larmor-drift mode
\begin{equation}
\omega^2=- \omega_{ci}^2 \frac{k_z^2}{k_\perp^2}\frac{T_e}{T_i}\frac{m_p}{m_e}\frac{\partial \ln n_e T_e}{\partial \ln n_i T_i}
\end{equation}
Here $\partial \ln A / \partial \ln B= (\partial \ln A/ \partial x) / (\partial \ln B/ \partial x) =
(B \partial A /\partial x) / (A \partial B /\partial x)$ notation is used.
It is clear that $\omega \ll \omega_{LD}$ and $\omega \ll \omega_{ci}$ conditions hold too (confirming in retrospect)
because  $k_z \to 0$. Equation (14) shows that condition for instability is
\begin{equation}
\frac{\partial \ln n_e T_e}{\partial \ln n_i T_i}>0
\end{equation}
which for uniform temperature plasma $T_e=T_i=const$ reduces to \\
${\partial \ln n_e} /{\partial \ln n_i} =1 >0$ is always fulfilled when
density variation for ions and electrons is the same.
Note that the plasma will be drift unstable for both  
increasing (positive) and decreasing (negative) density gradients because the ratio 
${\partial \ln n_e} /{\partial \ln n_i}$ will be {\it always positive} (negative/negative or positive/positive
is positive).
So, we conclude that our inhomogeneous plasma set up is always Larmor-drift, aperiodically
unstable.
It is also interesting to note that \cite{abr88}, p.169 showed that
aperiodic instabilities can lead to density filamentation (creating of
spatially thin threads). We can indeed see similar filamentary structures in density (and $E_x$) 
in Figures 1(a),1(b) and 3(a),3(c).

We gather from Figures 1(a) and 1(b), that $E_x$ and $n_e -n_{e0}$ perturbations
travel rather slowly compared to $E_y$ and the fast part of $B_z-B_{z0}$ (dark oblique strips in Figure 1(d)).
As can be inferred from Equation (12) and thick solid curve in Figure 7(c) (for $90 <x <95$), the lengthscale of the
background density gradient is $L_{IH} \approx 5 c/\omega_{pe}$ which 
roughly corresponds to the distance travelled by $E_x$ and $n_e -n_{e0}$ perturbations
(see e.g. start and end positions of rightmost bright streak in Figure  1(a) or location of the
rightmost peaks in Figures  3(a) and 3(c)), i.e. $11-6=5 c/\omega_{pe}$. This distance is travelled
in time $50 \omega_{pe}^{-1}$. Thus, the phase speed is estimated as $0.1 c$.
Generally $E_x$ and $n_e -n_{e0}$ perturbations repeat the shape of the density gradient, and other
runs (not shown here), with varied density gradient strength, confirm this.
$E_y$ perturbation as can be seen from Figure 1(c) travels from the density
gradient edges with a speed $\approx c$ (as the slope of dark and bright bands is unity).
$B_z-B_{z0}$ (Figures  1(d) and 3(d)) perturbation has two parts: slow moving part that
travels with speed $0.1c$ (as in  $E_x$ and $n_e -n_{e0}$) and and smaller, leading pulses
which travel with speed of $c$. We also gather from 3(f) that electron temperature is
slightly increased (i.e. the electron distribution function gets broader at 
$t=50 \omega_{pe}^{-1}$ (solid curve) compared to $t=0$ (dotted curve)).
  
To estimate frequency both of the slow and fast perturbations,
we note the number of bright features along e.g. 
left edge at $x=5$ in Figure  1(a) which is 7, counting from the first. 
Note that the density gradient left edge where $n_0 \approx 1$ is at $x=5$. 
The frequency estimate can be better done using Figure 2 where we plot $E_x(x=5,t)$ and $E_y(x=5,t)$.
The estimate is as follows.
In Figure 2(a) the time difference between leftmost and 
rightmost peaks is $\Delta t = 47.2-3=44.2 \omega_{pe}^{-1}$.
Thus
$7 (1/f) = 44.2 \omega_{pe}^{-1} = 44.2 \times (2\pi f_{pe})^{-1}$, and
$f \simeq f_{pe}$. Similar calculation for  Figure 2(b) yields
the time difference between leftmost and 
rightmost peaks is $\Delta t = 47.5-5=42.5 \omega_{pe}^{-1}$.
Thus
$7 (1/f) = 42.5 \omega_{pe}^{-1} = 42.5 \times (2\pi f_{pe})^{-1}$, and
$f=1.035f_{pe}$. We therefore conclude that 
$E_x$ oscillates at local plasma frequency and corresponds to
a plasma wave. Whereas $E_y$ perturbations are EM type (escaping radiation) and oscillate just above
the plasma frequency $1.035f_{pe}$.

We would like to stress that the generation of perturbations 
in the considered Larmor drift unstable case is {\it not}
due to the fact that pressure balance is not kept. In our case
plasma beta is small, therefore it is not crucial to keep thermodynamic pressure in balance and
the initial background density stays intact throughout the simulation time (see e.g. Figure 7(c), 
thick solid curve for $90 <x <95$).
We have performed numerical runs where temperature was varied as
inverse of $n_0(x)$ so that $p_0 = n_0(x)k_B T_0(x) = const$
($B_{z0}$  is constant throughout this study) 
yielding perfect total pressure balance -- similar approach to
pressure balance was adopted by \inlinecite{2005A&A...435.1105T}.
We confirm that even when total pressure balance was kept,
Larmor drift instability still developed and physical system 
behaviour was similar to what is presented here.

We note that if finite $B_{0y}=B_{0\perp}$  is added, this will change the Larmor-drift
stability criterion. This instability will stabilise if \cite{1964SvPhU...7..209R}:
\begin{equation}
\frac{B_{0\perp}}{B_{z0}} \geq \frac{v_{th,i}}{\omega_{c i} L_{IH}}.
\end{equation}
We plan to study this stabilisation issue in a following publication
(work in progress), using EPOCH 1.5D particle-in-cell code which allows
to choose all background magnetic field components (not as VALIS,
which only allows to consider $(E_x,E_y,0)$ and $(0,0,B_z)$). 

We conclude this subsection with the observation that we found a new possibility 
for exciting plasma frequency EM radiation by means of a universal, aperiodic
Larmor-drift instability. By universal we mean that satisfying the 
instability criterion (see Equation (14)) is quite plausible in many astrophysical situations.
Condition $\omega \ll \omega_{ci}$ requires that 
${k_z^2}/{k_\perp^2} \simeq {L_\perp^2}/{L_z^2} \ll {m_e}/{m_p}$
i.e. length of the domain should be at least $\approx 43$ times longer than its width.

\subsection{Plasma emission case, homogeneous plasma with low density beam}

To suppress the Larmor-drift instability (because we cannot achieve this by imposing suitable
$B_{0x}$) 
we now set uniform normalised plasma  
number density to $n_0=1$ and inject a low density beam with the parameters specified above. 

\begin{figure}    
   \centerline{\hspace*{0.015\textwidth}
               \includegraphics[width=0.515\textwidth,clip=]{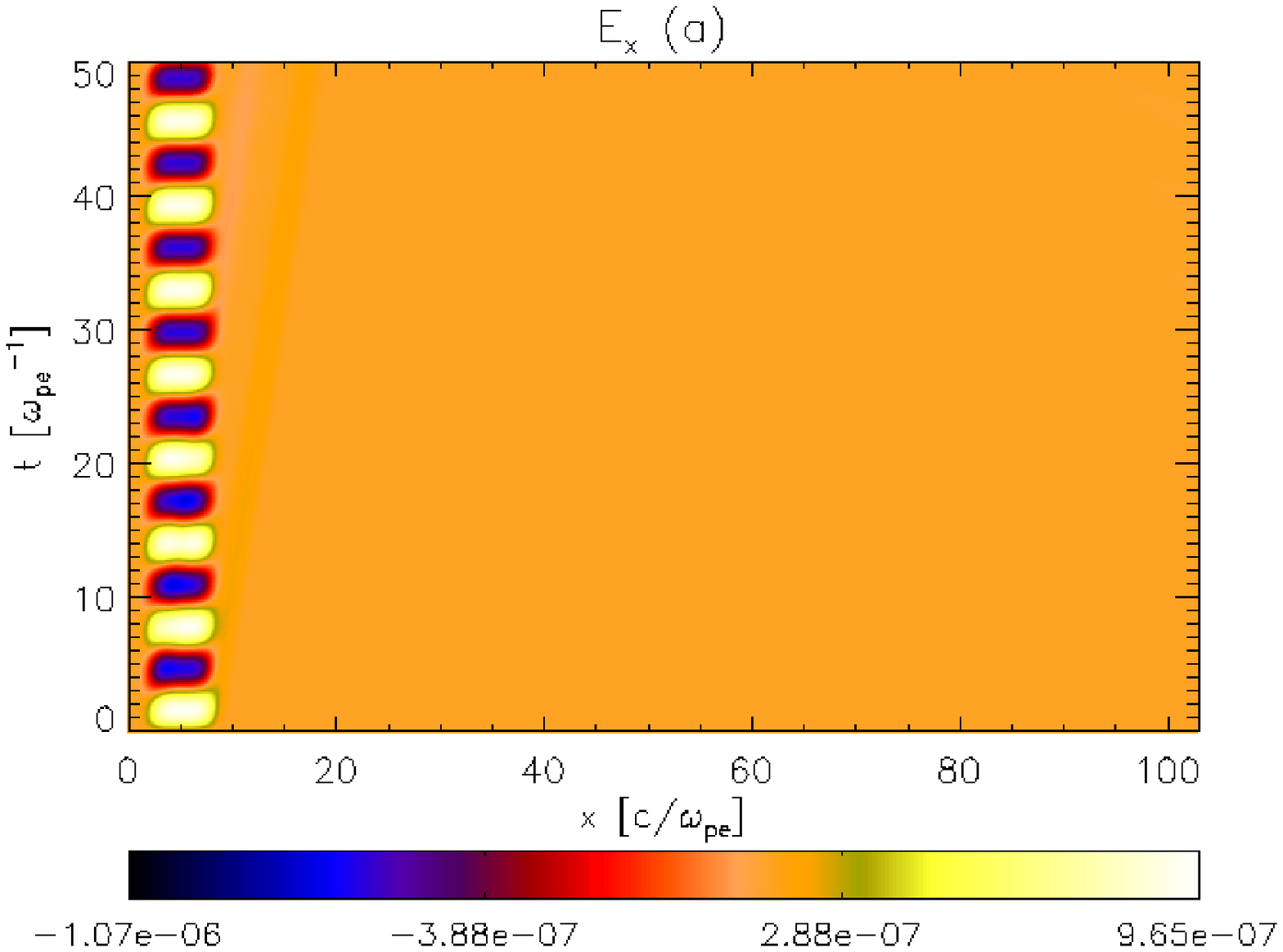}
               \hspace*{-0.03\textwidth}
               \includegraphics[width=0.515\textwidth,clip=]{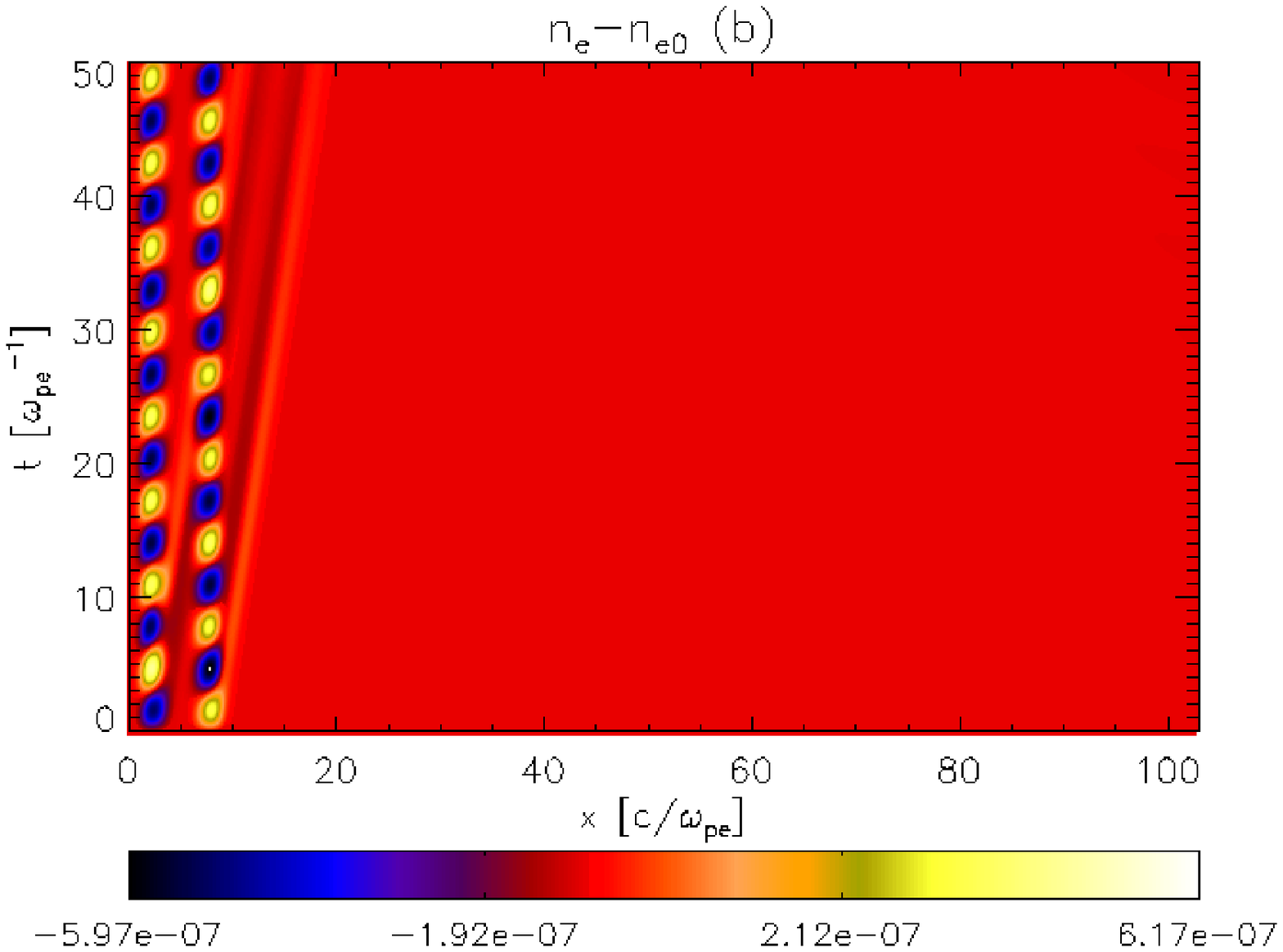}
              }
   \centerline{\hspace*{0.015\textwidth}
               \includegraphics[width=0.515\textwidth,clip=]{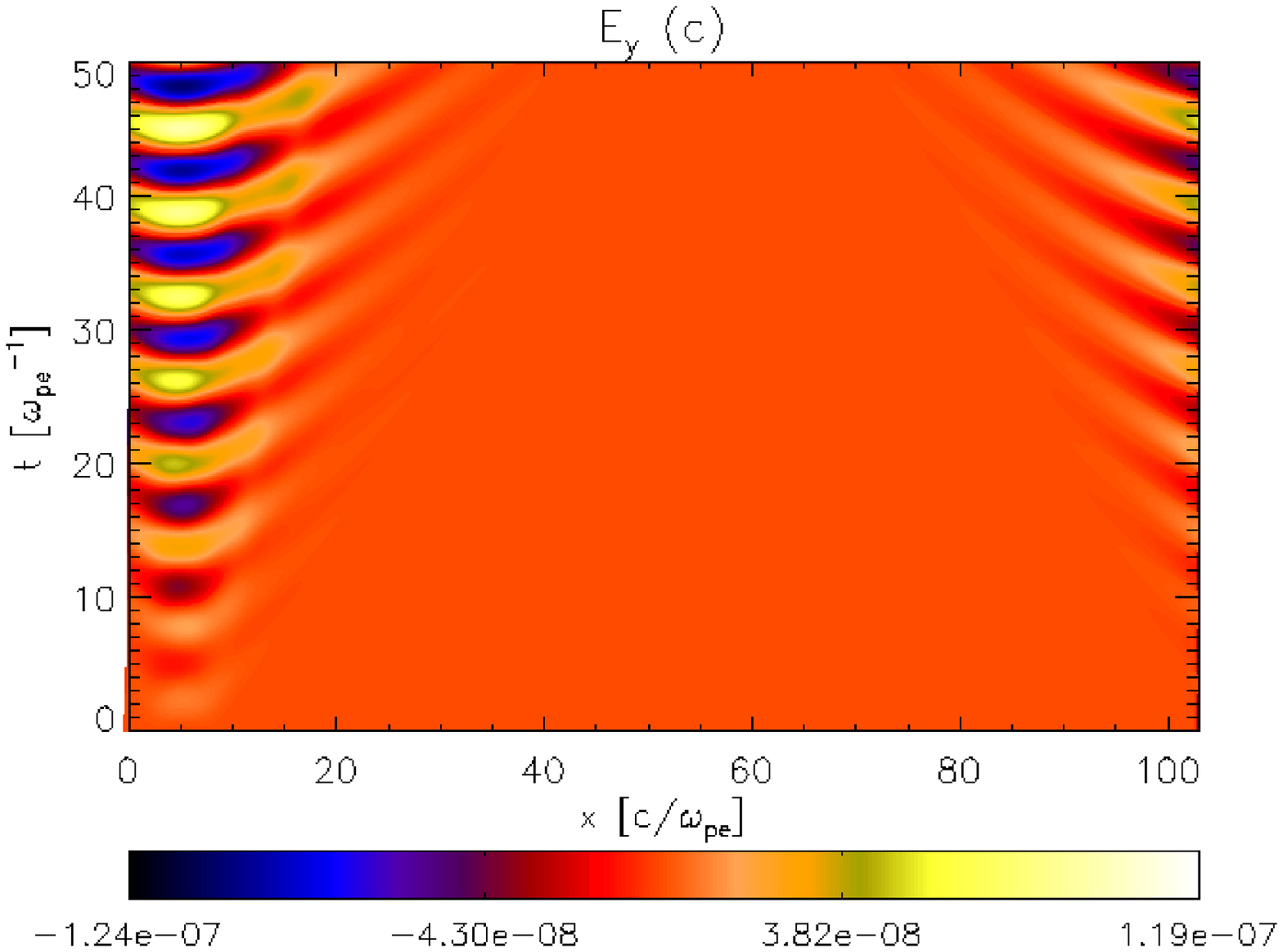}
               \hspace*{-0.03\textwidth}
               \includegraphics[width=0.515\textwidth,clip=]{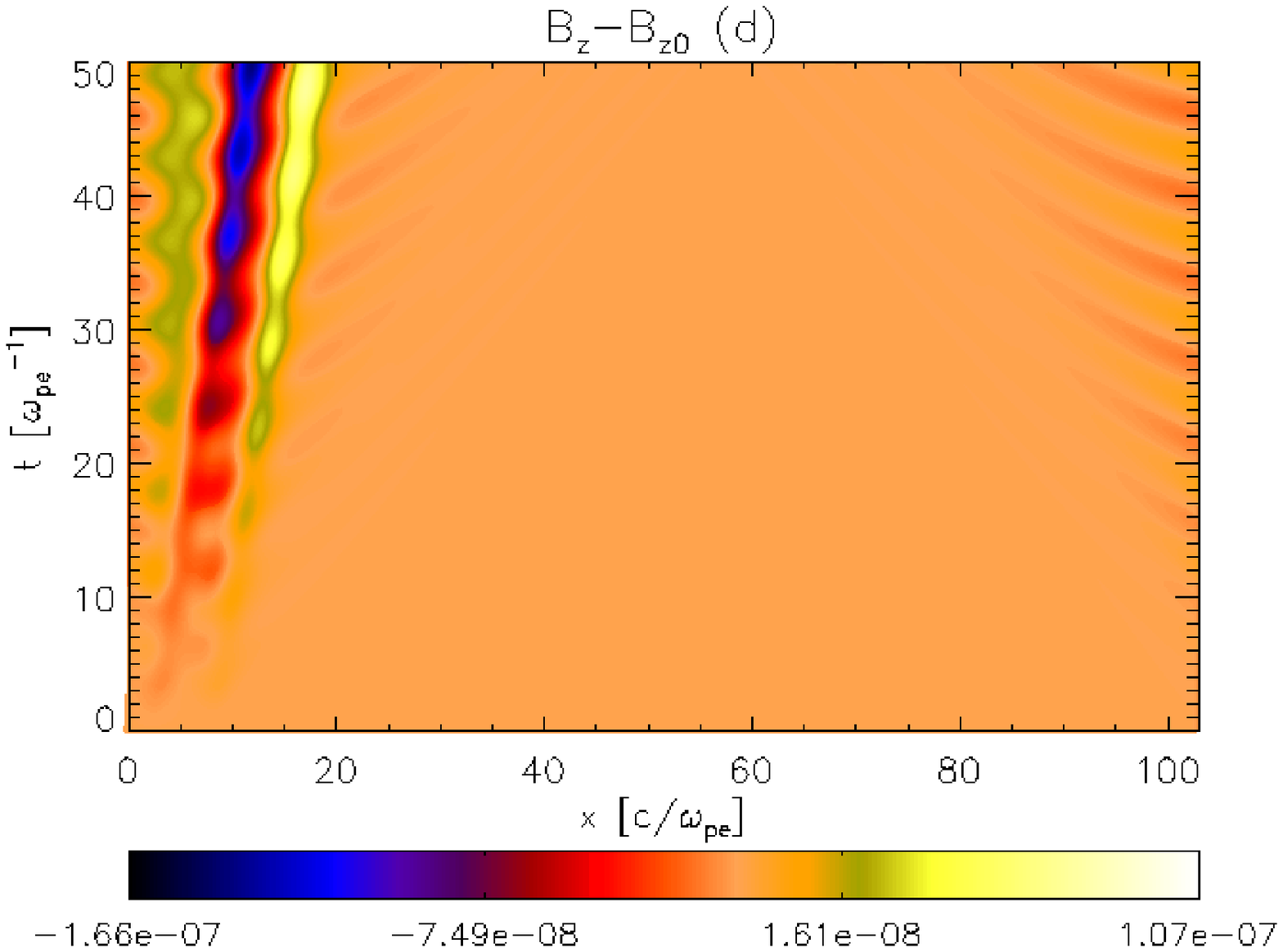}
              }
\caption{As in Figure  1 but for the case of homogeneous plasma with low density beam.}
     \end{figure}
   
 \begin{figure}    
   \centerline{\includegraphics[width=0.99\textwidth,clip=]{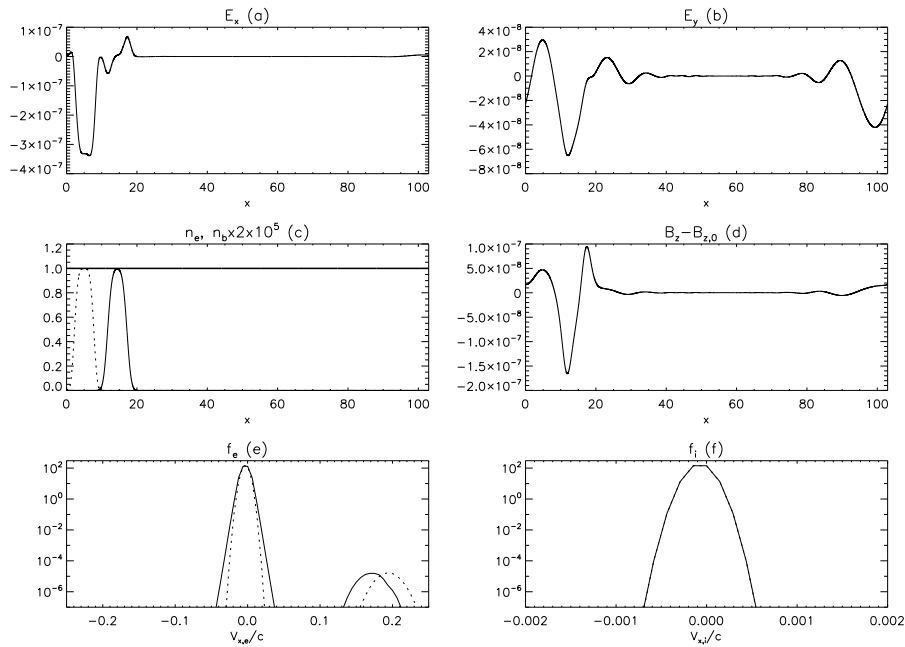}
              }
              \caption{(a) $E_x(x,t=50)$, 
 (b) $E_y(x,t=50)$,  (c) $n_{e}(x,t=50)$ (thick solid horizontal line), 
$n_b(x,t=0)$ (dotted curve) and $n_b(x,t=50)$ (thin solid curve),
(note that $n_b$ was scaled by a factor of $2\times10^5$ to make it visible),  (d) $B_z(x,t=50)-B_{z0}$, 
 (e) $f_e(v_x,t=0)$ (dotted curve) and $f_e(v_x,t=50)$  (solid curve), 
 (f) $f_i(v_x,t=50)$  (solid curve) ($f_i(v_x,t=0)$ is also plotted with a dotted curve, but
to a plotting accuracy the curves overlap, indicating no ion heating takes place). }
     \end{figure}

The results of this numerical run are presented in Figures 4 and 5,
while time dynamics is presented in movie 2 (see the accompanying electronic supplementary material).
One striking novel feature immediately seen in Figures 4(a) and 4(b)
($E_x$ and $n_e -n_{e0}$ where $n$'s include initially injected beam contribution)
is the excitation of standing waves in the spatial location of
the beam injection. By counting the number of bright features in the
elapsed time, it is clear that the oscillations are at about $\omega_{pe}$.
Moreover, $E_x$ oscillates as one solid feature in the spatial location of
beam injection (oscillation spatial 
width coincides with the beam width). While $n_e -n_{e0}$  also exhibits standing waves, but
these are in the regions of positive and negative density gradients
of the back and front of the  {\it beam} (recall that here background plasma is homogeneous).
These oscillations are in anti-phase, i.e. at given $t=const$
over-density and under-density is observed.
As with Larmor-drift instability (Section 2.1) this was unforeseen result.
However, again 
literature analysis enabled us to find a suitable interpretation.
There are two possible regimes. If the phase of the waves is locked, in the 
strong instability regime,
waves can appear with the frequency close to $\omega_{pe}$ in the location of the
beam injection \cite{pp77}. In the case of a weak turbulence regime, the formation of
strong Langmuir waves also observed near the beam injection
site \cite{1999SoPh..184..353M} (according to their Equation (15)
Langmuir turbulence spectral energy density 
near the beam injection point depends on the phase velocity as $W\propto v^5$).
In the electromagnetic $E_y$ and $B_z-B_{z0}$ components (Figures  4(c) and 4(d))
again standing wave at the beam injection location can be also seen, but in addition
this serves as a source to the escaping EM radiation. These can be seen
as oblique dark and bright strips with a slope close to unity, thus propagating at 
about speed of light. 
Wide dark oblique strip (with narrow bright front) in Figure 4(d) with a slope
$\Delta x / \Delta t = (15-5)/50 = 0.2$ corresponds to the beam wake
(recall that beam travels with speed $0.2c$).
Note that in  4(c) and 4(d) there are also EM waves present near $x\approx 100$.
This is because the standing wave centred on $x=5$ generates EM waves travelling in 
both directions. Because of the periodic boundary conditions, waves that travel to the
left, appear on the right side of the simulation domain. 

Figure  5 provides further details: 5(a) shows $E_x$ at time $t=50 \omega_{pe}^{-1}$
and is made of two parts (i) a deep centred on $x=5$ corresponds to the
standing wave at plasma frequency and it nicely coincides with 
beam injection site, see dotted pulse in 5(c) which shows the beam at $t=0$;
(ii) a smaller hump in 5(a) at $x\approx 16$ which roughly coincides
with the location to where beam has travelled in  time $t=50 \omega_{pe}^{-1}$
(solid pulse in Figure 5(c) centred at $x\approx 15$).
Figure 5(e) indicates that bulk plasma electron distribution function
heats up (solid peak centred on $v_e=0$
which corresponds to $t=50 \omega_{pe}^{-1}$ is wider than at $t=0$).
Also we can see that the beam slows down from $0.2c$ to $0.17c$ 
(small bump (dotted curve) centred at $0.2c$ which represents the beam at $t=0$
shifts to $0.17c$ (solid curve bump) which is the beam at $t=50 \omega_{pe}^{-1}$).
We note that since the beam is nine times hotter than the background plasma, $T_b=9T=9.0\times10^5$K, 
Larmor radius of the beam is three times larger than that of background plasma, $r_{L,b}= 1.24 c/\omega_{pe}$. 
This is smaller than the distance traveled by the beam ($\approx 10 c/\omega_{pe}$ cf. Figure 5(c)). 
Thus, the beam partial magnetisation can be regarded as the main cause of its slowing down. 
We also observe that there is no substantial quasilinear relaxation
(i.e. plateau does not form) which corroborates basic features
of the quasilinear theory. This is due to the fact that  
the growth rate of resonant 
Langmuir waves given by Equation (7) is small, as in this run ${n_b}/n_e=5 \times 10^{-6}$.
A simpler estimate for quasilinear relaxation time, $\tau$, 
(time of establishing the plateau) is 
achieved by
using $\tau =n_e/({n_b} \omega_{pe})$ (e.g. \cite{1999SoPh..184..353M}).
Based on this, we see that in our case $\tau=2\times10^5 \omega_{pe}^{-1}$.
Thus, it is not surprising that in 50 plasma frequencies we do not see substantial
quasilinear relaxation.
\inlinecite{1999SoPh..184..353M} quote the criterion of weak turbulence regime
of quasilinear theory to apply as
$\varepsilon \equiv n_b m_e v_b^2 / (n_0 m_e v_{th,e}^2) \ll 1$.
Here, $\varepsilon \approx 10^{-2} \ll 1$, thus we are well in the quasilinear  regime.
Another interesting corroboration of the quasilinear-theory is that 
shape of the beam does not change i.e. despite the fact that 
small density
beam plunges through the background plasma at a speed of $0.2c$, it stays
intact. \inlinecite{1999SoPh..184..353M} offer suitable explanation for this fact
based on the beam particle kinematics. To avoid duplication we refer the interested 
reader to this paper for the details.
Figure 5(f) confirms that despite the fact ions are treated in the numerical
code as moving, still there is no significant change in their velocity distribution function.

It is interesting to note that newly established standing waves 
can offer an alternative explanation for the horizontal strips observed in some dynamical spectra.
\inlinecite{2010A&A...515A...1A} report a narrow-band, short duration 
line emission at 314 MHz which is interpreted as a gyro-resonance line.
We note that the observed feature can also be explained by an EM emission
emanating from the standing waves. This naturally explains the fact that
there is no drift in frequency, as the standing wave {\it remains} in the same
spatial location, hence there is no change in density and in emission
(plasma) frequency. Moreover, if we look at the dynamical spectra from Figure 1 from 
\inlinecite{2010A&A...515A...1A} we see that initially, the line intensity increases in time.
This behaviour is very similar to what is seen our 4(c) where the intensity increase
at $x=5$ in time can be seen.
We can roughly estimate intensity of the line predicted by our model as follows.
From Figures  5(b) and 5(d) we can read off typical amplitudes of the escaping
EM radiation in the form of standing wave as $E_y \simeq 10^{-7}$ and $B_z'=B_z-B_{z0} \simeq 2\times 10^{-7}$.
Flux of the Poynting vector is than $F=|\vec F| \simeq E_y B_z'/\mu_0$. Recovering 
physical units from the normalised quantities yields $F \approx 5\times10^{-4}$ Wm$^{-2}$.
Assuming the area of the emitting source is $A=$20 Mm$\times$1 Mm and plasma frequency of $f_p=300$ MHz,
we can estimate the flux density at 1 AU as $F_d = F A/[4 \pi (1AU)^2 f_p]\approx 1.2 \times 10^{-22}$ Wm$^{-2}$ Hz$^{-1}$ 
($\approx 1$ sfu). This is of the order of the \inlinecite{2010A&A...515A...1A} estimate of few sfu for the
line flux density.

In summary, the plasma emission case in a homogeneous plasma with low density beam
confirms the general picture of quasilinear theory even in the case when the beam is
injected {\it transversely} to the magnetic field, with a main novelty being
established  that the spatial location where beam is injected 
serves as a source for standing plasma waves which, in turn, generate escaping
EM radiation that
oscillats at plasma frequency. This offers a possible new interpretation of 
the horizontal strips observed in some dynamical spectra.
     
\subsection{Larmor drift-unstable, plasma emission case, inhomogeneous plasma with dense beam}

We now combine both effects considered in Sections 2.1 and 2.2 by considering
Larmor drift-unstable plasma emission case in which  plasma is inhomogeneous and  dense beam
is injected transverse to the magnetic field and along the density gradient.
\begin{figure}    
   \centerline{\hspace*{0.015\textwidth}
               \includegraphics[width=0.515\textwidth,clip=]{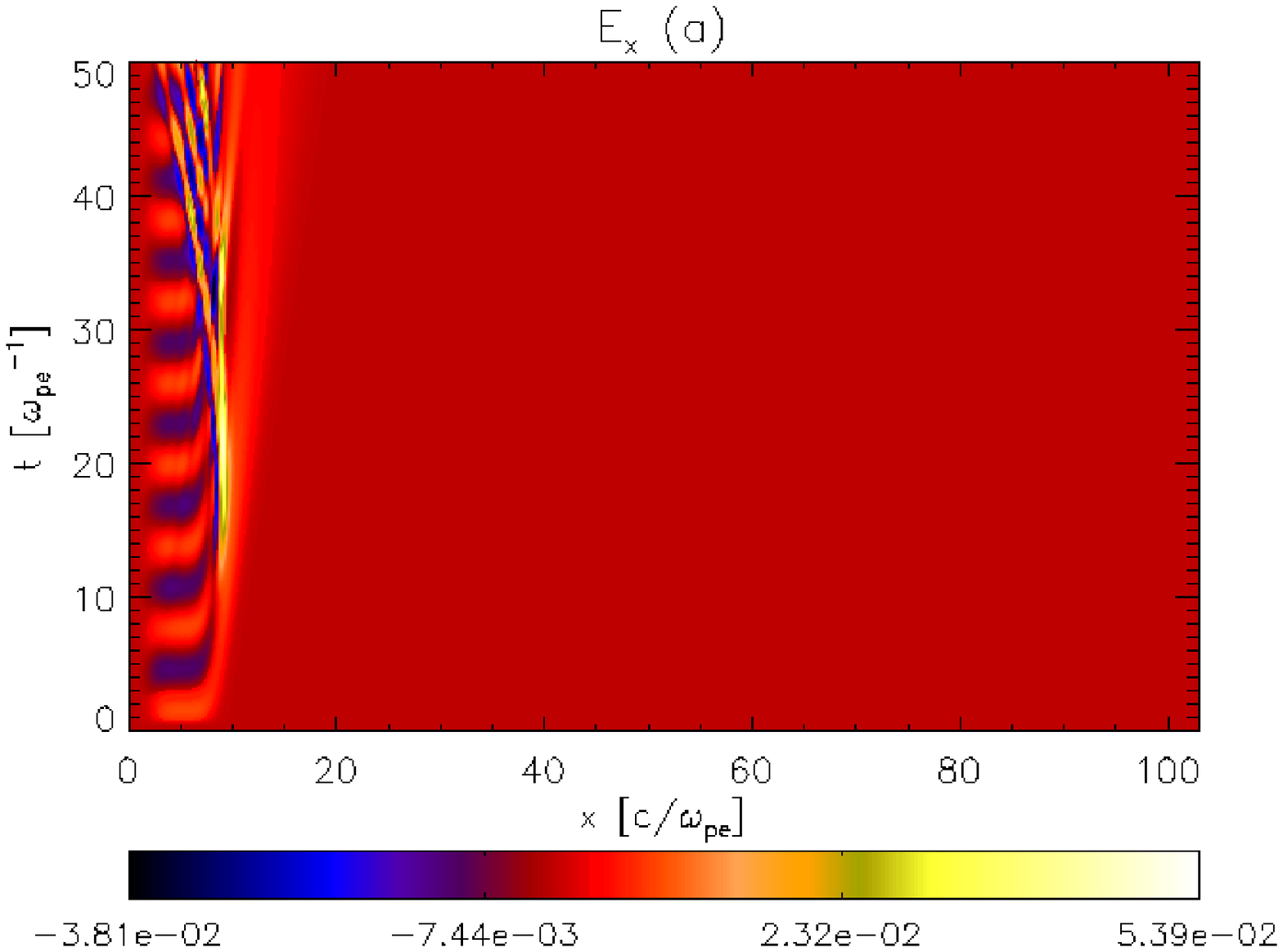}
               \hspace*{-0.03\textwidth}
               \includegraphics[width=0.515\textwidth,clip=]{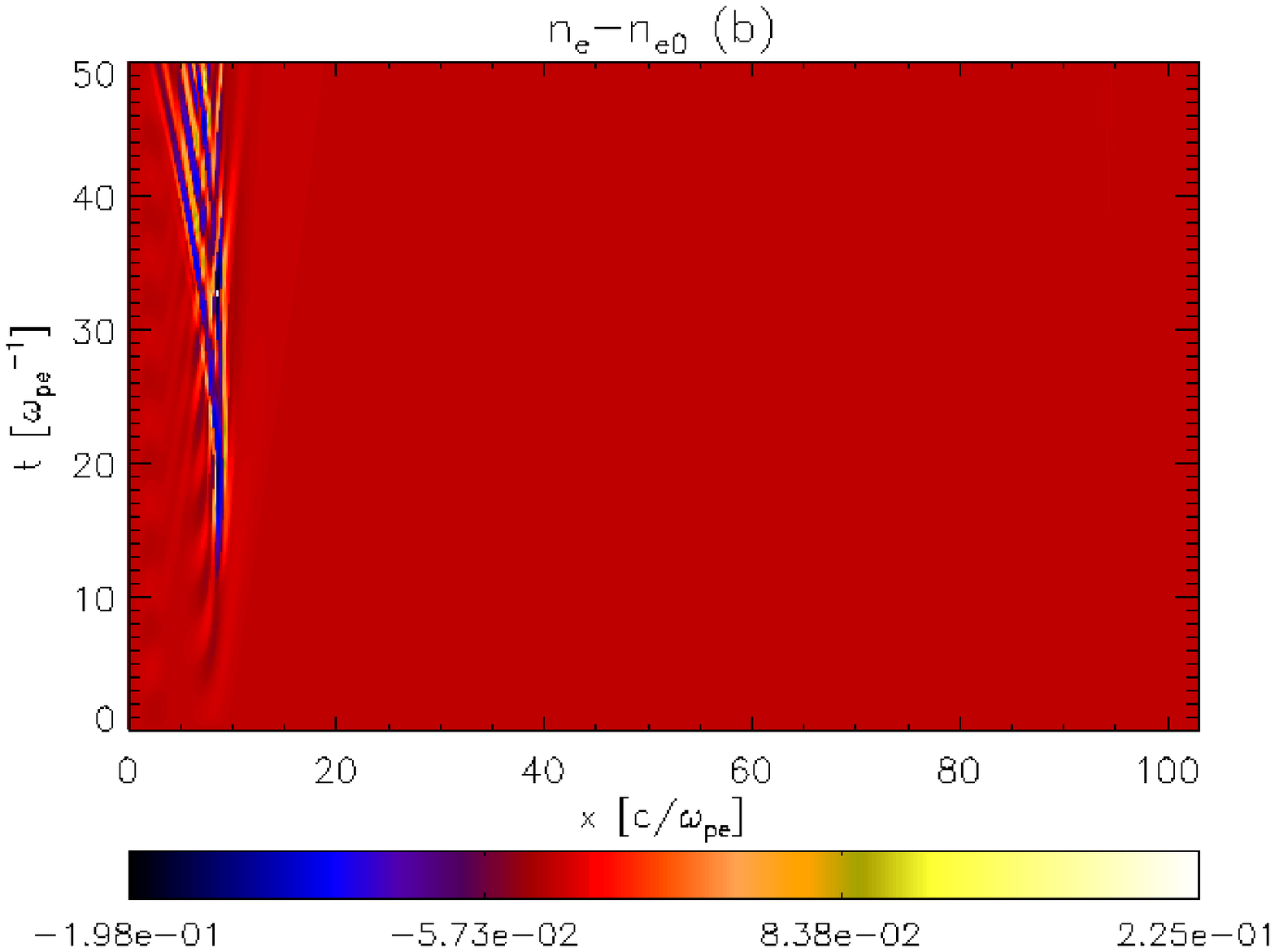}
              }
   \centerline{\hspace*{0.015\textwidth}
               \includegraphics[width=0.515\textwidth,clip=]{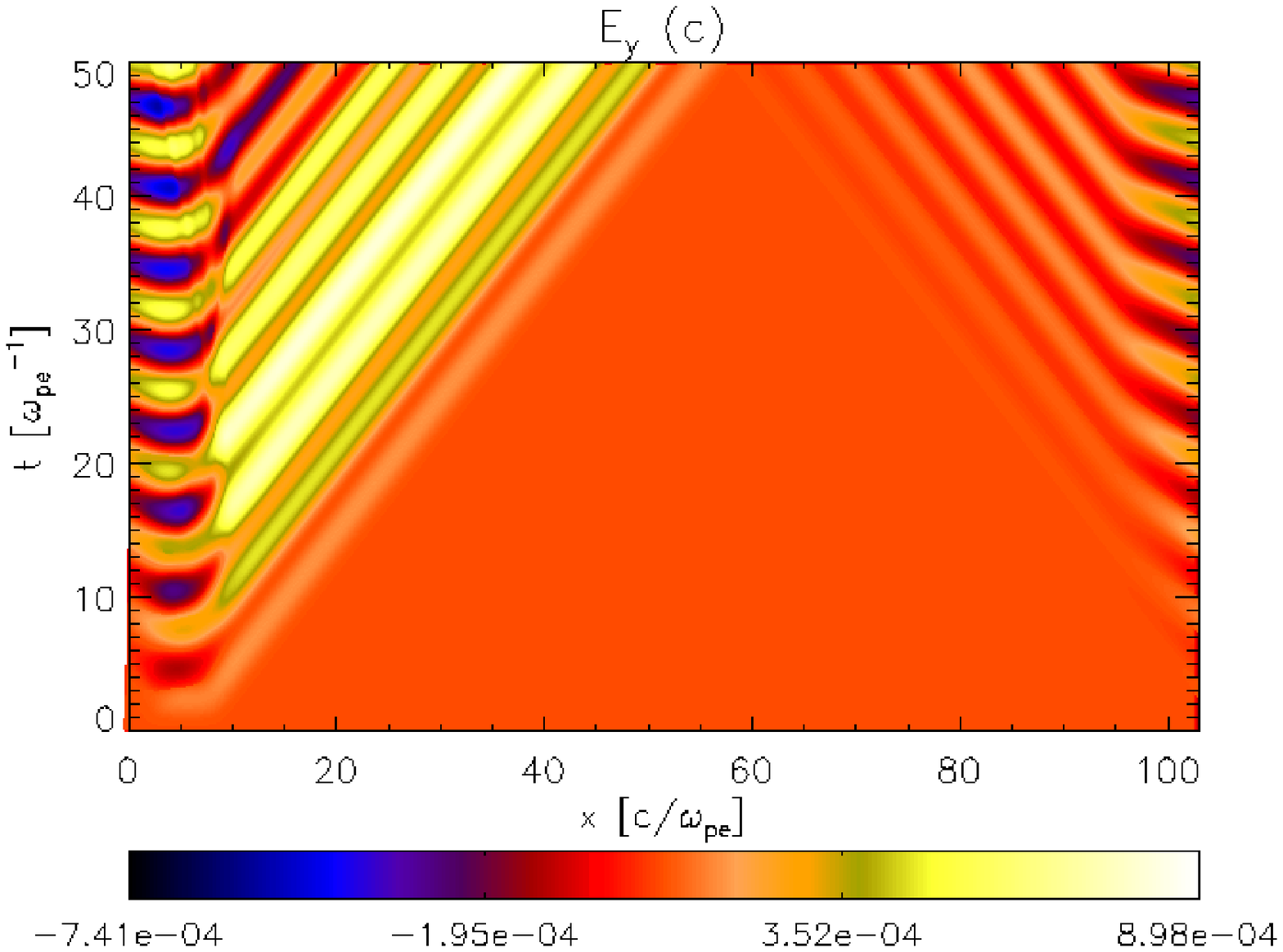}
               \hspace*{-0.03\textwidth}
               \includegraphics[width=0.515\textwidth,clip=]{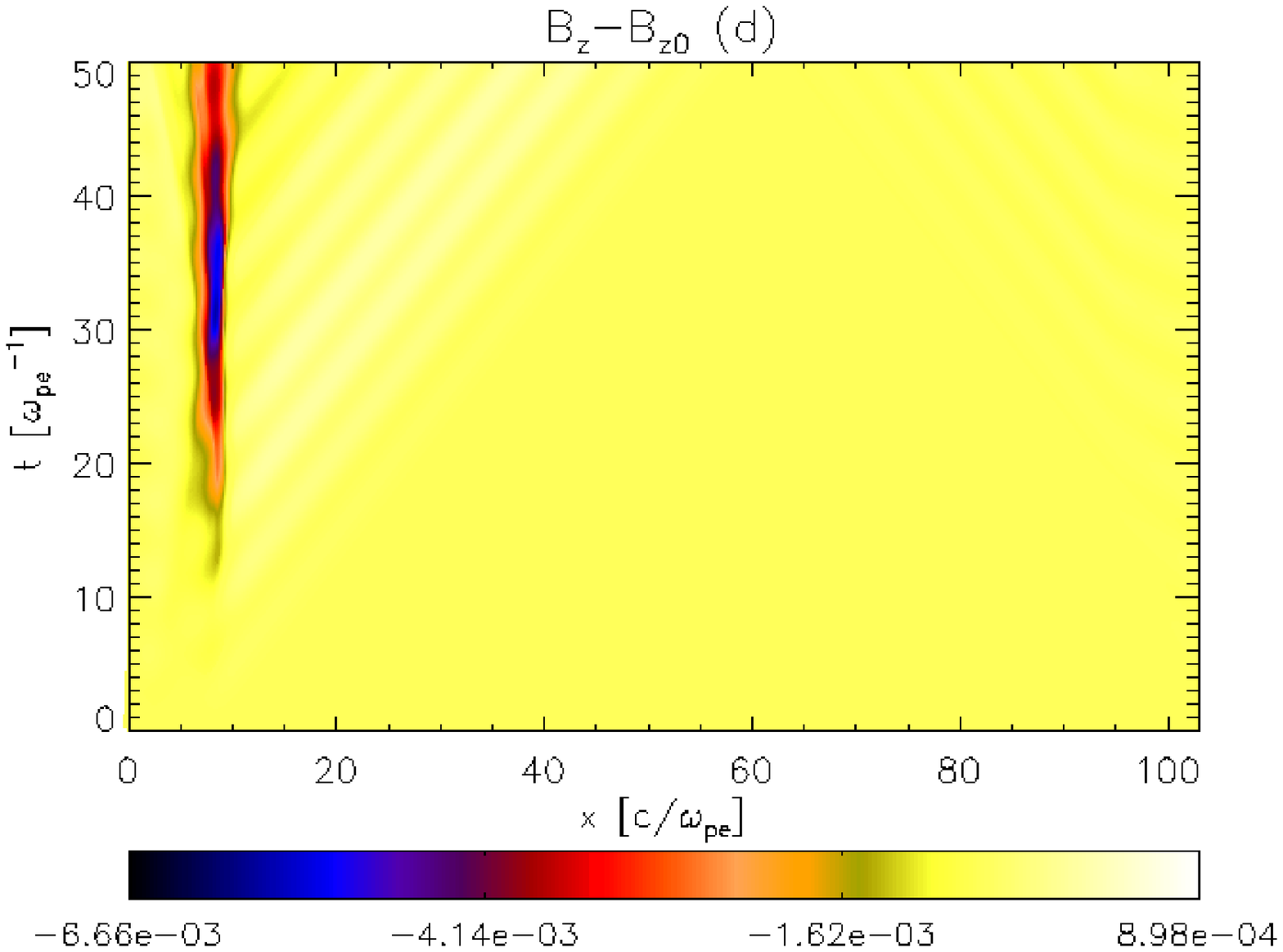}
              }
\caption{As in Figure  1 but for the case of inhomogeneous plasma with dense beam.}
      \end{figure}
   
 \begin{figure}    
   \centerline{\includegraphics[width=0.99\textwidth,clip=]{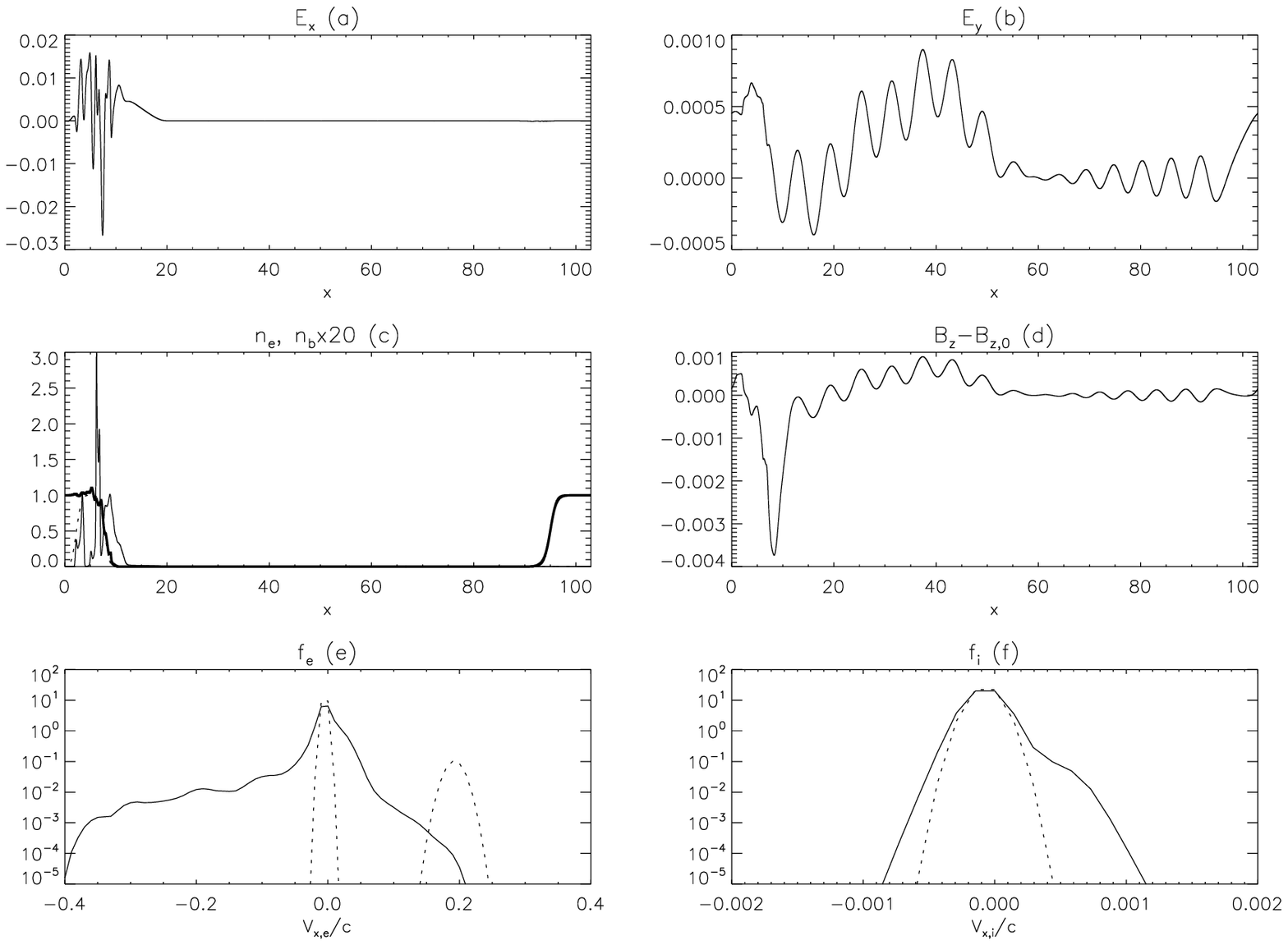}
              }
              \caption{(a) $E_x(x,t=50)$, 
 (b) $E_y(x,t=50)$,  (c) $n_{e}(x,t=50)$ (thick solid curve), 
$n_b(x,t=0)$ (dotted curve) and $n_b(x,t=50)$ (thin solid curve),
(note that $n_b$ was scaled by a factor of $20$ to make it visible),  (d) $B_z(x,t=50)-B_{z0}$, 
 (e) $f_e(v_x,t=0)$ (dotted curve) and $f_e(v_x,t=50)$  (solid curve), 
 (f)  $f_i(v_x,t=0)$ (dotted curve) and $f_i(v_x,t=50)$  (solid curve).  }
      \end{figure}

We gather from Figure 6(a) that $E_x$ (plasma wave component) now comprises of 
two parts: a weak standing wave at the location of the beam injection which oscillates
with frequency $\omega_{pe}$ and a strong wake created by the beam which now shows 
much more dispersion, as we depart from the quasilinear regime where 
non-linear interactions between wave modes are ignored.
Time-distance plot for $n_e -n_{e0}$ (Figure 6(b) is dominated by a wake created by the beam).
EM wave components ($E_y$ and $B_z-B_{z0}$) as in Section 2.2 show similar behaviour
where the standing wave centred on $x=5$ generates escaping EM waves travelling in 
both directions. 

Figure 7(a) shows that $E_x$ has two parts: a small leftmost bump which is from a
standing plasma wave and a beam generated perturbation which travels roughly with the beam speed.
Figures  7(b) and 7(d) for $E_y$ and $B_z-B_{z0}$, show in more detail, how 
the standing wave centred on $x=5$ generates EM waves travelling in 
both directions with speed of light.
We gather from Figure 7(c) that as the dense beam plunges into the plasma,
it no longer retains its shape as in quasilinear regime (compare to Figure  5(c)).
Figures 7(e) and 7(f) show dynamics of electron and ion distribution functions.
We see that as the beam/background plasma number density ratio, 
${n_b}/n_e=5 \times 10^{-2}$, is no longer small, two effects can be observed:
(i) quasilinear relaxation happens very fast; and (ii) substantial
electron return current (wide wing with negative velocities in Figure 7(e)) is generated.
For ions sizable heating takes place (Figure 7(f)).
Dynamical picture of the system evolution can be studied using movie 3 (see the accompanying electronic supplementary material).
In the considered case $\tau=20 \omega_{pe}^{-1}$.
Thus, it is not surprising that in 50 $\omega_{pe}^{-1}$ we see substantial
quasilinear relaxation taking place.
The criterion of weak turbulence regime
$\varepsilon \equiv n_b m_e v_b^2 / (n_0 m_e v_{th,e}^2) \ll 1$ is no longer fulfilled as
here $\varepsilon \approx 10^2 \gg 1$. Thus the physical system is not 
in the quasilinear regime.

In summary, we see for this set of results that the system is driven by the effects
of the beam while Larmor-drift effect, whilst present, is not dominant.
Significant deviation from the quasilinear theory is found which
manifests itself in (i) fast quasilinear relaxation, (ii) 
disintegration of the beam, and (iii) generation of significant electron return 
current and ion heating.

\section{Conclusions}

We used 1.5D Vlasov-Maxwell simulations to model EM emission generation in a fully 
kinetic model for the first time in the solar physics context. We studied 
plasma emission mechanism and Larmor drift instability
in a single plasma thread that joins the Sun to Earth with 
the spatial scales compressed appropriately. 
The results can be summarised in three points:
\begin{itemize}
\item We established that 1.5D inhomogeneous plasma with a uniform background 
magnetic field directed transverse to the density
gradient is aperiodically unstable to Larmor-drift instability. 
This instability 
results in a novel effect of generation of EM emission at plasma frequency. The 
generated perturbations
consist of two parts: (i) non-escaping (trapped) Langmuir type oscillations which are
are localised in the regions of density inhomogeneity, are highly filamentary, and the period of 
appearance of the filaments is close to
electron plasma frequency in the dense regions; and (ii) escaping electromagnetic radiation with phase 
speeds close to the
speed of light. 
\item In the uniform density plasma case (when plasma becomes stable to Larmor drift instability),
when a {\it low density} super-thermal, hot beam is injected along the domain,
in the direction perpendicular to the magnetic field,
plasma emission mechanism generates non-escaping Langmuir type {\it standing} oscillations
at the beam injection location,
which in turn generate escaping electromagnetic radiation at the electron plasma frequency.
This result can be used to offer an alternative interpretation to
the horizontal strips (usually referred to as 
the narrowband emission lines in the literature) 
observed in some dynamical spectra. 
Quasilinear theory predictions: (i) electron free streaming and (ii) long relaxation
time, in accord with the analytic expressions, are confirmed via direct,
fully-kinetic simulation.
\item We considered interplay of Larmor-drift instability and plasma emission mechanism
by studying a {\it dense} electron beam in the Larmor-drift unstable (inhomogeneous) plasma. 
We established that in this case the physical system is driven by the effects
of the beam while Larmor-drift is not dominant. We
also found significant deviation from the quasilinear theory which
manifests itself in (i) fast quasilinear relaxation, (ii) 
disintegration of the beam, and (iii) generation of significant electron return 
current and ion heating.
\end{itemize}

We would like to close with the comments in relation to the prospects of comparison of the 
numerical simulations presented here with the observations (e.g. with the 
dynamical spectra -- 2D radio emission intensity plots 
where frequency is on $y$-axis and time on $x$-axis).
Let us base our argument on the level of generated $E_y$ (EM component) in the
considered three cases (Sections 2.1-2.3). In the Larmor-drift unstable case
without the electron beam according to Figure 1(c) $E_y$ attains values around 
$5\times 10^{-7}$.
In the uniform (Larmor-drift stable case) with a low density beam (see. Figure 4(c)),
$E_y$ attains only $10^{-7}$ (five times less).
In the Larmor-drift unstable case with dense beam $E_y$ attains $9\times 10^{-4}$ (Figure 6(c)).
Based on this we conclude at this stage we cannot produce numerical (synthetic)
dynamical spectrum where frequency of the radiation would drop in time as the 
beam moves along the decreasing density profile. This is for two reasons
(i) when we consider small density beam ${n_b}/n_e=5 \times 10^{-6}$
(when we are in the quasilinear regime and all known facts about plasma emission
mechanism apply) {\it the beam effect is too small} ($E_y=10^{-7}$)
compared to the Larmor-drift unstable case without the electron beam ($E_y=5\times 10^{-7}$),
thus there is no point presenting results of Larmor-drift unstable case with weak beam
(because Larmor-drift instability overwhelms the effect of the weak beam).
(ii) In the Larmor-drift unstable case with dense beam (${n_b}/n_e=5 \times 10^{-2}$),
 we have $E_y=9\times 10^{-4}$, and
we do not see decrease of the emission frequency with time because
the beam disintegrates in quasilinear time of $\tau = 20 \omega_{pe}^{-1}$
(i.e. beam has not enough time to slide down the decreasing density profile)
and hence we cannot expect to see plasma emission mechanism in action
in its usual form. 
In summary, the model presented here cannot be 
directly applied to the type III busts and it is more relevant for the 
interpretation of the narrowband line emission observations.  
In oder to simulate the dynamical spectra of type III bursts
in which the emission intensity rapidly drifts towards small frequencies in time, as
the beam moves to a plasma with decreasing density (and hence $\omega_{pe}$), we would first need
to suppress the Larmor-drift instability. In turn,
this can be achieved by satisfying the condition set out in Equation (16).
However, with the VALIS numerical code this is not possible because it only solves for 
$(E_x,E_y,0)$ and $(0,0,B_z)$. We plan to study this issue further in a following publication
(work in progress), using EPOCH 1.5D particle-in-cell code which allows
to choose all background magnetic field components.
Naturally, PIC method will suffer from the known shortcomings compared to the superior Vlasov-Maxwell approach. However, the benefit of
ability of specifying all EM field components outweighs the downsides of the PIC approach. 

\begin{acks}
The author would like to thank T.D. Arber (Warwick) for useful discussions about VALIS numerical
code. Also, useful discussions with E. Kontar (Glasgow) and H. Aurass (AIP) at the CESRA2010 conference
are gratefully acknowledged. Computational facilities used are that of Astronomy Unit, 
Queen Mary University of London and STFC-funded UKMHD consortium at St Andrews
University. The author would like to thank HEFCE-funded 
South East Physics Network (SEPNET) for financial support.
\end{acks}

\end{article} 
\end{document}